\documentclass[twocolumn,final]{elsarticle}
\usepackage{graphicx}
\usepackage{braket}

\journal{Journal of Molecular Spectroscopy}

\newcommand*{\wn}{cm$^{-1}$}
\newcommand*{\BX}{$B^1\Sigma^+_u-X^1\Sigma^+_g$}
\newcommand*{\CX}{$C^1\Pi_u-X^1\Sigma^+_g$}
\newcommand*{\EFX}{$EF^1\Sigma_g^+-X^1\Sigma^+_g$}
\newcommand*{\EF}{$EF^1\Sigma_g^+$}
\newcommand*{\GK}{$GK^1\Sigma_g^+$}
\newcommand*{\X}{$X^1\Sigma^+_g$}
\newcommand*{\Hm}{H$_{2}$}
\newcommand*{\Dm}{D$_{2}$}

\newcommand*{\HDi}{HD$^{+}$}
\newcommand*{\Htwoi}{H$_{2}^{+}$}

\def\apb{Appl.\  Phys.\ B}
\def\apjl{Astroph.\  J.\  Lett.\ }
\def\apj{Astroph.\  J.\ }

\def\jcp{J. Chem.\ Phys.\ }
\def\cjp{Can.\ J.\ Phys.\ }

\def\pra{Phys.\ Rev.\ A }
\def\prd{Phys.\ Rev.\ D }
\def\prl{Phys.\ Rev.\ Lett.\ }
\def\pr{Phys.\ Rev.\  }
\def\plb{Phys.\ Lett.\ B }

\def\nph{Nat.\ Phys.\ }
\def\npb{Nucl.\ Phys.\ B }
\def\cpl{Chem.\ Phys.\ Lett.\ }
\def\jms{J. Mol.\ Spectrosc.\ }
\def\jpb{J. Phys.\ B }

\def\molp{Mol.\ Phys.\ }

\def\mnras{Mon.\ Not.\ Roy.\ Astron.\ Soc.\ }
\def\pccp{Phys.\ Chem.\ Phys.\ Chem.\ }

\def\aap{Astron.\ Astrophys.\  }
\def\eujpd{Eur.\ J. \ Phys. \ D }

\begin{document}

\begin{frontmatter}

\title{Physics beyond the Standard Model from hydrogen spectroscopy}

\author{W. Ubachs\corref{mycorrespondingauthor}}
\cortext[mycorrespondingauthor]{Corresponding author}
\ead{w.m.g.ubachs@vu.nl}
\author{J. C. J. Koelemeij}
\author{K. S. E. Eikema}
\author{E. J. Salumbides}
\address{Department of Physics and Astronomy, Vrije Universiteit, De Boelelaan 1081, 1081 HV Amsterdam, The Netherlands}

\begin{abstract}
Spectroscopy of hydrogen can be used for a search into physics beyond the Standard Model. Differences between the absorption spectra of the Lyman and Werner bands of H$_2$ as observed at high redshift and those measured in the laboratory can be interpreted in terms of possible variations of the proton-electron mass ratio $\mu=m_p/m_e$ over cosmological history. Investigation of some ten of such absorbers in the redshift range $z= 2.0-4.2$ yields a constraint of $|\Delta\mu/\mu|< 5 \times 10^{-6}$ at 3$\sigma$. Observation of H$_2$ from the photospheres of white dwarf stars inside our Galaxy delivers a constraint of similar magnitude on a dependence of $\mu$ on a gravitational potential $10^4$ times as strong as on the Earth's surface.
While such astronomical studies aim at finding quintessence in an indirect manner, laboratory precision measurements target such additional quantum fields in a direct manner.
Laser-based precision measurements of dissociation energies, vibrational splittings and rotational level energies in H$_2$ molecules and their deuterated isotopomers HD and D$_2$ produce values for the rovibrational binding energies fully consistent with quantum ab initio calculations including relativistic and quantum electrodynamical (QED) effects. Similarly, precision measurements of high-overtone vibrational transitions of HD$^+$ ions, captured in ion traps and sympathetically cooled to mK temperatures, also result in transition frequencies fully consistent with calculations including QED corrections. Precision measurements of inter-Rydberg transitions in H$_2$ can be extrapolated to yield accurate values for level splittings in the H$_2^+$-ion.
These comprehensive results of laboratory precision measurements on neutral and ionic hydrogen molecules can be interpreted to set bounds on the existence of possible fifth forces and of higher dimensions, phenomena describing physics beyond the Standard Model.
\end{abstract}


\end{frontmatter}


\section{Introduction}

With the detection and first characterization of the Higgs particle~\cite{Atlas2012,CMS2012} the Standard Model of physics has seemingly become complete.
A full description in three families of fundamental material particles behaving as fermions, acting upon each other through three fundamental forces via bosonic force-carrying particles, has been formulated in a consistent framework.
However, a number of physical phenomena are not contained in the Standard Model (SM): dark matter and dark energy are not explained; a connection of the three fundamental SM-forces to gravity cannot be made; and neither is it understood why gravity is such a feeble force. Deeper questions as to why the Universe is built in 3+1 dimensions, why there exist only 4 forces, and whether the constants of nature may depend on cosmological history or on local space-time conditions remain unanswered as of yet.

It is the paradigm of the present paper that molecular physics may contribute to finding answers to these questions. The recent searches for a dipole moment of the electron, setting tight constraints of $|d_e| < 8.7 \times 10^{-29}$ e$\cdot$cm, have already ruled out some supersymmetric extensions to the Standard Model. These studies take advantage of the very strong internal polarizing fields of $\sim 10^{11}$ V/cm in YbF~\cite{Hudson2011} and ThO~\cite{ACME2014} molecules as well as of HfF$^+$ ions~\cite{Loh2013}.

At the most fundamental level laser spectroscopic investigations of molecules can be applied for an experimental test of the symmetrization postulate of quantum mechanics. The fact that integer-spin species must exhibit a symmetric wave function upon interchange of identical particles makes that certain rotational quantum states in the $^{16}$O$_2$~\cite{Deangelis1996,Naus1997,Gianfrani1999} or in the $^{12}$C$^{16}$O$_2$~\cite{Modugno1998,Mazotti2001} molecule cannot exist. This provides an experimental platform for the investigation of the symmetrization postulate.

A search for varying constants implies a search for physics beyond the SM. The fact that the fundamental coupling strengths and other constants such as the mass ratios between particles are inserted as parameters in the fundamental theory suggests that these fundamental constants can be varied at will. However, a time-variation or a space-time dependence of these constants should be phrased in field theories, and in order for those to remain consistent, even in the simplest approaches, additional fields have to be invoked~\cite{Bekenstein2002,Barrow2002a}. Such fields can also be associated with fifth forces or quintessence~\cite{Carroll1998}.

The hydrogen molecule is an experimental probe for detecting a possible variation of the proton-electron mass ratio $\mu=m_p/m_e$, a dimensionless constant of nature. The strong dipole-allowed spectrum of the Lyman and Werner bands in the vacuum ultraviolet spectral range provides a sensitive tool for detecting H$_2$.
Observation of H$_2$ absorption spectra, redshifted into the atmospheric transmission window ($\lambda > 300$ nm) with ground-based telescopes constrain putative variations of $\mu$ over time.
Since H$_2$ is the most abundant molecule in the Universe, it is readily observed even up to redshifts as high as $z=4.2$~\cite{Bagdonaite2015} providing look-back times of 12.5 billion years into cosmic history.
Alternatively, H$_2$ absorption spectra as observed from photospheres of white dwarf stars can be employed to probe whether this constant $\mu$ depends on special local conditions, such as e.g. gravitational fields.

Another approach toward probing new physics from molecules is to carry out ultra-precise measurements on level energies of molecules under well-controlled laboratory conditions. A historical inspiration is drawn from the discovery of the Lamb shift in atomic hydrogen~\cite{Lamb1947}, where a measurement of a small shift in the relative energies between $2s_{1/2}$ and $2p_{1/2}$ levels led to the birth of quantum electrodynamics (QED), the modern theory describing the interaction between light and matter.
In a similar fashion QED-calculations of small molecular systems can be tested in precision laser spectroscopic experiments to search for phenomena beyond QED.
Any deviations from first principles calculations in the framework of QED can be interpreted in terms of physics beyond the Standard Model, in terms of fifth forces or quintessence, in terms of extra dimensions, or as any other phenomenon beyond the Standard Model. Here the status of such searches for new physics from spectroscopy of hydrogen molecules and molecular ions will be discussed.

The remainder of the paper is structured as follows. In sections~\ref{muvar} and \ref{mu-grav} the use of the H$_2$ spectrum for a search for a varying proton-electron mass ratio is discussed, either on a cosmological time scale or as a local variability with dependence on a gravitational field. In section~\ref{H2metrology} a number of recent precision measurements on the H$_2$ molecule, including its dissociation limit, its fundamental ground state vibrational splitting, as well as level energies of a sequence of rotational states and a high vibrational level, are presented. In section~\ref{H2ionmetrology} some recent precision measurements on the H$_2^+$/HD$^+$ molecular ions are discussed.
These outcomes of experimental precision studies are compared with the most advanced QED-theoretical calculations for these benchmark molecules in section~\ref{QED-theory}. The agreement found between the experiments and theory is interpreted, in section~\ref{BSM-physics}, in terms of setting bounds on a fifth force and on extra dimensions, both concepts beyond the Standard Model of physics. In the final section~\ref{outlook} an outlook is presented on perspectives to perform more sensitive searches for new physics in these areas.

\section{A cosmologically varying proton-electron mass ratio}
\label{muvar}

Searches for temporal variations of the  proton-electron mass ratio $\mu=m_p/m_e$ can be conducted by comparing absorption spectra of molecules observed astronomically from high redshift objects with the same spectra measured in the laboratory. Such comparisons can be made for many molecules~\cite{Jansen2014}, but molecular hydrogen is often a species of choice since it is the most abundant molecule in the Universe, and readily observed at medium to high redshifts. The relative variation of $\mu$ can be deduced by imposing the following relation to the obtained redshifted wavelengths $\lambda^z_i$ and laboratory (zero redshift) wavelengths $\lambda^{\rm{0}}_i$:
\begin{equation}
 \frac{\lambda^z_i}{\lambda^{\rm{0}}_i} = (1+z_{\rm{abs}}) \left( 1+K_i\frac{\Delta\mu}{\mu} \right).
\label{comp-equation}
\end{equation}
Here the redshift parameter $z_{\rm{abs}}$ relates to the overall redshift of the absorbing hydrogen cloud, and the parameters $K_i$ represent the sensitivity coefficients expressing the induced wavelength shift upon a varying $\mu$ for each individual line in the absorption spectrum of H$_2$~\cite{Ubachs2007}:
\begin{equation}
  K_i = \frac{d\ln \lambda_i}{d \ln \mu}.
\label{sens-coeff}
\end{equation}

The laboratory wavelengths $\lambda^{\rm{0}}_i$ are determined in laser-based experiments of the \BX\ Lyman and \CX\ Werner bands of H$_2$ via extreme ultraviolet laser spectroscopy~\cite{Philip2003,Ubachs2004,Reinhold2006,Hollenstein2006} achieving accuracies of $\Delta\lambda/\lambda =5 \times 10^{-7}$. Even better accuracies were obtained in a study measuring anchor level energies in two-photon Doppler-free excitation of the \EFX\ system~\cite{Hannemann2006}, combined with
results of a comprehensive Fourier-transform emission study delivering relative values for rovibrational levels in ${EF}^1\Sigma^+_g$, $GK^1\Sigma^+_g$, $H^1\Sigma^+_g$, $B^1\Sigma^+_u$, $C^1\Pi_u$, $B'^1\Sigma^+_u$, $D^1\Pi_u$, $I^1\Pi_g$, and $J^1\Delta_g$ states~\cite{Bailly2010}.
This two-step process yielded accuracies of $\Delta\lambda/\lambda =5 \times 10^{-8}$~\cite{Salumbides2008} for a number of Lyman and Werner band lines. For the purpose of comparing with quasar absorption spectra the laboratory values for the absorption wavelengths may be considered exact.

The overall redshift $z_{\rm{abs}}$ is, within a good approximation, connected to a look-back time $T$ into the cosmological history of the Universe
\begin{equation}
\label{look-back}
 T = T_0 \left[ 1 - \frac{1}{(1+z_{\rm{abs}})^{3/2}} \right]
\end{equation}
where $T_0 = 13.8$ Gyrs is the age of the Universe.

\begin{figure}
\resizebox{1\linewidth}{!}{\includegraphics{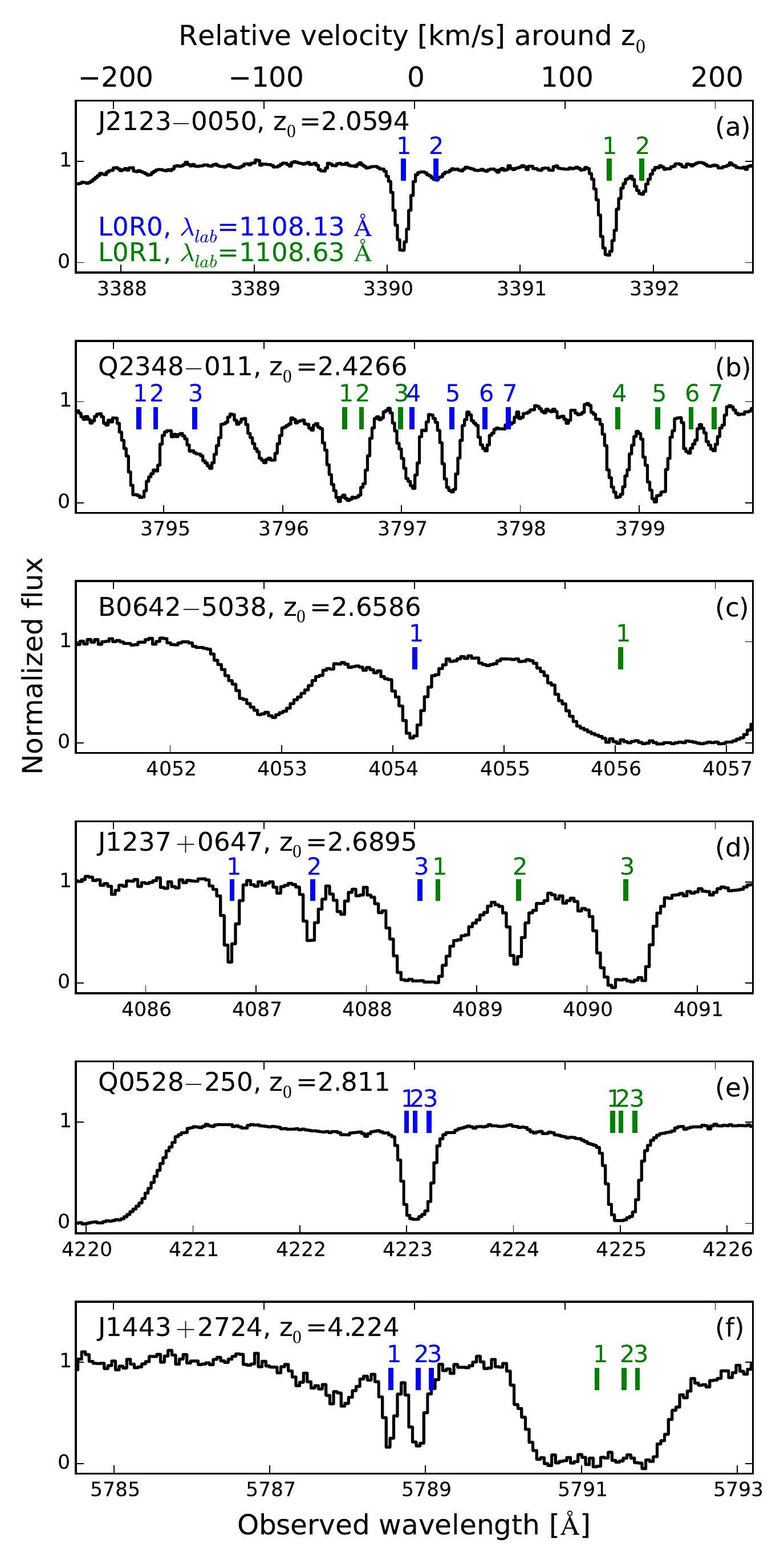}}
\caption{Observed spectra toward six different quasars with molecular hydrogen absorbing systems at redshifts in the range $z=2.05 - 4.2$. The part of the spectrum for rest wavelengths 1106 - 1109 \AA\ covering the R(0) and R(1) lines in the \BX\ (0,0) Lyman band are shown, appearing at varying longer wavelengths in high-redshift absorption profiles. Spectra are also plotted on a velocity scale (upper scale in km/s) with respect to a chosen redshift $z_0$ as indicated.}
\label{Quasarspec}
\end{figure}

A possible small differential effect on the absorption spectrum, due to a varying proton-electron mass ratio and represented by a relative change $\Delta\mu/\mu$, may be derived from the spectrum in a comprehensive fit. In such a fit the molecular physics information on the wavelengths $\lambda^0_i$, the oscillator strength $f_i$, and the natural lifetime broadening parameters $\Gamma_i$ for each H$_2$ line is fixed, the values for the sensitivity coefficients $K_i$ inserted, and the physical parameters describing the absorbing medium, such as the Doppler width $b$, the column densities for each $J$-level $N_J$ and the redshift $z_{\rm{abs}}$ are fitted to model the spectrum. From the resulting values of $N_J$ it is found that typical population temperatures of ~50 K are found in most cases, although the populations do not exactly represent a Boltzmann distribution. In all cases only absorption from the lowest $v''=0$ vibrational level is observed for rotational states $J=0-5$.
Hence the dipole allowed absorption spectrum lies shortward of 1120 \AA\ and extends to the Lyman cutoff of 912 \AA, thus covering the \BX\ Lyman bands from (0,0) up to (17,0) and the \CX\ Werner bands up to (4,0).
In several cases absorption lines of deuterated hydrogen HD are observed and included in the analysis, by comparing to accurate laboratory wavelengths~\cite{Ivanov2008a} and invoking calculated sensitivity coefficients~\cite{Ivanov2010}. For HD only R(0) lines are observed.
Due to the opacity of the Earth's atmosphere (transparency for $\lambda > 300$ nm) only for systems of redshift $z>2$ a significant number of absorption lines can be observed with ground based telescopes.

In Fig.~\ref{Quasarspec} six different H$_2$ absorption spectra at redshifts in the range $z=2.0-4.2$ are displayed as observed in the line of sight of quasars.
Each system displays a number of distinct velocity features, ranging from a single feature as in B0642-5038 to seven distinct features as in Q2348-011. In Fig.~\ref{Quasarspec}  information about quasar data is compiled from original analyses for quasars J2123-0050 at $z=2.0594$~\cite{Malec2010,Weerdenburg2011}, for Q2348-011 at $z=4.266$~\cite{Bagdonaite2012}, for B0642-5038 at $z=2.6586$~\cite{Bagdonaite2014a}, for J1237+0647 at $z=2.6895$~\cite{Dapra2015}, Q0528-250 at $z=2.811$~\cite{King2011}, and for J1443+2724 at $z=4.224$~\cite{Bagdonaite2015}.
In some panels, blends with typical Lyman-$\alpha$ forest lines are seen, like for the R(1) line in B0642-5038 and in J1443+2724. Such features have to be included in the fitting procedure. In a final analysis of each individual spectrum the parameter $\Delta\mu/\mu$ is included as a fitting parameter and determined. The uncertainty in the relative variation of this fundamental constant is obtained from the covariance matrix. After having performed the comprehensive fitting analysis of the six systems presented in Fig.~\ref{Quasarspec}, and the additional works on systems HE0027-1836~\cite{Rahmani2013}, Q0347-383~\cite{King2008,Thompson2009,Wendt2012}, Q0405-443~\cite{King2008,Thompson2009}, and Q1232+082~\cite{Varshalovich2001} an overall average is obtained of $\Delta\mu/\mu = (3.1 \pm 1.6) \times 10^{-6}$ in the redshift interval $z=2.0-4.2$, so for look-back times of 10.5-12.4 Gyrs. This finding of a positive value for $\Delta\mu/\mu$ at the 1.9$\sigma$ significance level is too small to be interpreted as evidence for a variation of a fundamental constant, and is interpreted as a null result: a constraint on a variation of the proton-electron mass ratio of
$|\Delta\mu/\mu| < 5 \times 10^{-6}$ (at 3$\sigma$).

\section{A gravitational dependence of the proton-electron mass ratio}
\label{mu-grav}

The spectrum of molecular hydrogen can also be employed to probe a possible dependence of the proton-electron mass ratio $\mu$ on environmental
conditions, connecting to the chameleon scenario for varying constants~\cite{Khoury2004}. In the photosphere near the surface of collapsed stars, also known as white dwarfs, the gravitational field is some 10$^4$ times stronger than at the Earth's surface. Hence the absorption spectra of H$_2$, in rare cases observed from such photospheres~\cite{Xu2013}, transfer a fingerprint of the value of $\mu$ existing in extreme gravitational field conditions~\cite{Bagdonaite2014b}.

\begin{figure}[ht!]
\resizebox{1\linewidth}{!}{\includegraphics{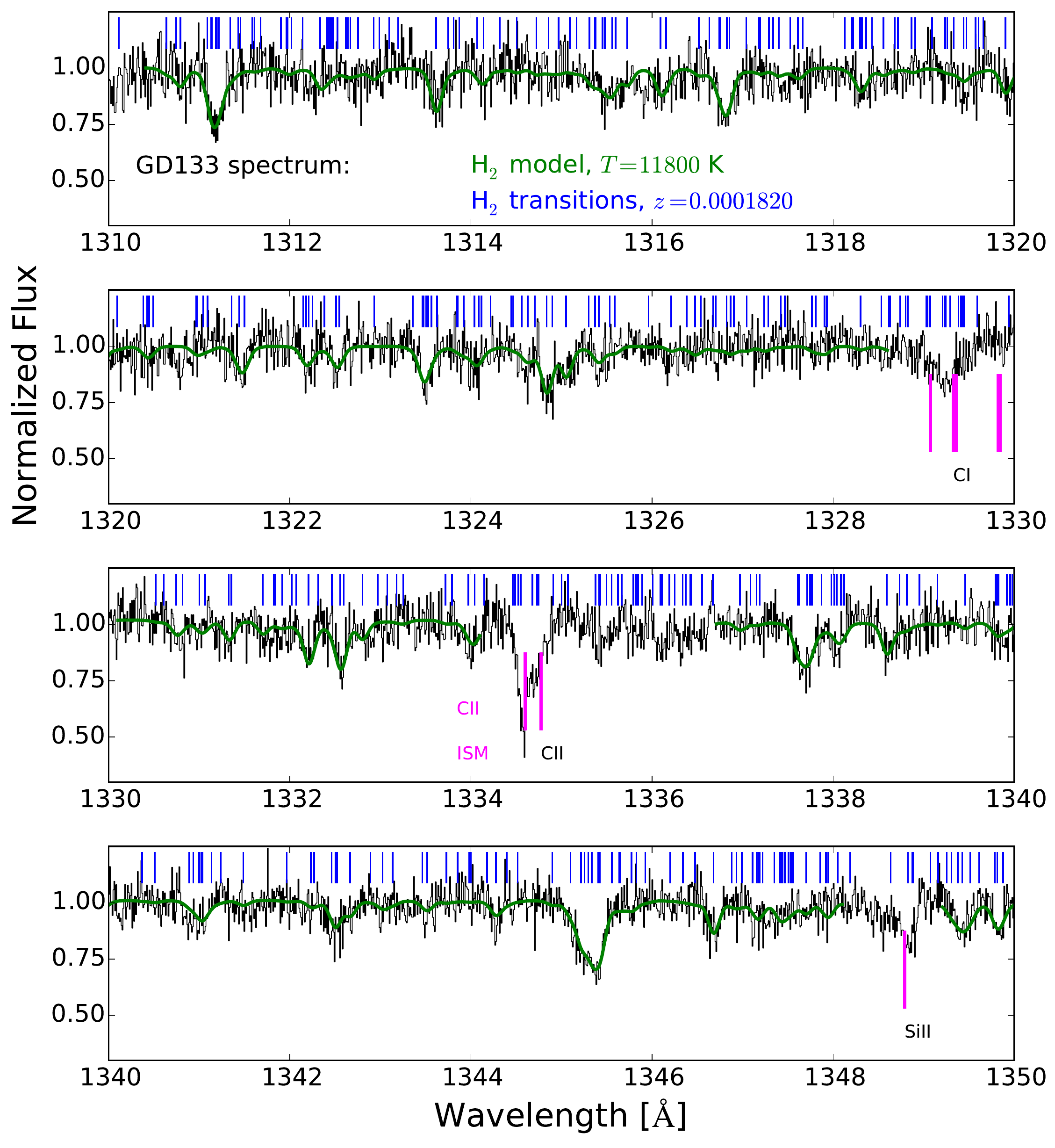}}
\caption{Part of the absorption spectrum of the white dwarf GD-133 as obtained with the Cosmic Origins Spectrograph aboard the Hubble Space Telescope in the range 1310-1350 \AA. The solid (green) line is the result of a fit with optimized parameters; the parts without the solid line were not included in the fit. The tick marks represent the positions of H$_2$ absorption lines contributing to the spectrum. }
\label{gd133}
\end{figure}

Fig.~\ref{gd133} displays part of the absorption spectrum of the white dwarf GD133 as observed with the Cosmic Origins Spectrograph aboard the Hubble Space Telescope. The spectrum consists of a large number of blended absorption lines in the range 1310-1410 \AA, which can be assigned to absorptions in Lyman bands \BX\ $(v',v'')$ bearing the largest intensities in bands (0,3), (0,4), and (1,3). The spectrum of this white dwarf is observed in our Galaxy, and is therefore not affected by a cosmological redshift, but the gravitational potential induces a gravitational redshift to the spectrum of $z_G=1/\sqrt{1-(2GM/Rc^2)}-1$, where $G$ is Newton's constant, $c$ the speed of light, and $M$ and $R$ the mass and the radius of the white dwarf star.

The experimental spectrum is compared to a spectral model function $S(\lambda)$ for H$_2$ described as:
\begin{equation}
\label{wd-spectrum}
 S(\lambda) = N_c \sum_{v'',J''} P_{v'',J''}(T) f_{v',J',v'',J''}(\lambda)
\end{equation}
with $N_c$ a total column density of absorbing H$_2$, $P_{v'',J''}(T)$ the partition function representing the quantum state populations as a function of temperature and including a 3:1 ortho-para distribution, and $f_{v',J',v'',J''}$ the oscillator strengths for individual transitions, which were calculated with \emph{ab initio} methods~\cite{Salumbides2015}. This spectral function was implemented into a fitting routine, where for each spectral line a Lorentzian width $\Gamma_i$ corresponding to the excited state lifetime was convolved with a Doppler width $b$. In the fit $N_c$, $b$, the temperature $T$, the gravitational redshift $z_G$ and $\Delta\mu/\mu$ were optimized. A good reproduction of the GD-133 spectrum was obtained for a temperature $T=11100 \pm 470$ K, a Doppler shift $b= 14.5 \pm 0.6$ km/s, a column density $\log N_c = 15.817 \pm 0.007$, and a redshift $z_G=0.0001819\,(11)$~\cite{Salumbides2015}. The high temperature explains why the populated vibrational levels $v''=3$ and $4$ contribute to the spectrum and why so many H$_2$ absorption lines lend intensity to the spectrum. In the modeling of the GD-133 spectrum over a thousand H$_2$ lines are included, represented by the sticks in Fig.~\ref{gd133}.

The comprehensive fit of the GD-133 white dwarf spectrum returns a value of $\Delta\mu/\mu = (-2.3 \pm 4.7) \times 10^{-5}$ delivering a constraint on a dependence of the proton-electron mass ratio on a gravitational potential of $\phi_{GD-133}=1.2 \times 10^{-4}$, which may be compared with the gravitational potential at the Earth's surface of $\phi_E=9.8 \times 10^{-9}$. Note that here the dimensionless gravitational potential is defined as $\phi=GM/Rc^2$. An analysis has been performed for another white dwarf GD29-38 yielding a similar result~\cite{Bagdonaite2014b,Salumbides2015}.

\section{Laser precision metrology of H$_2$}
\label{H2metrology}

\begin{figure}
\resizebox{1\linewidth}{!}{\includegraphics{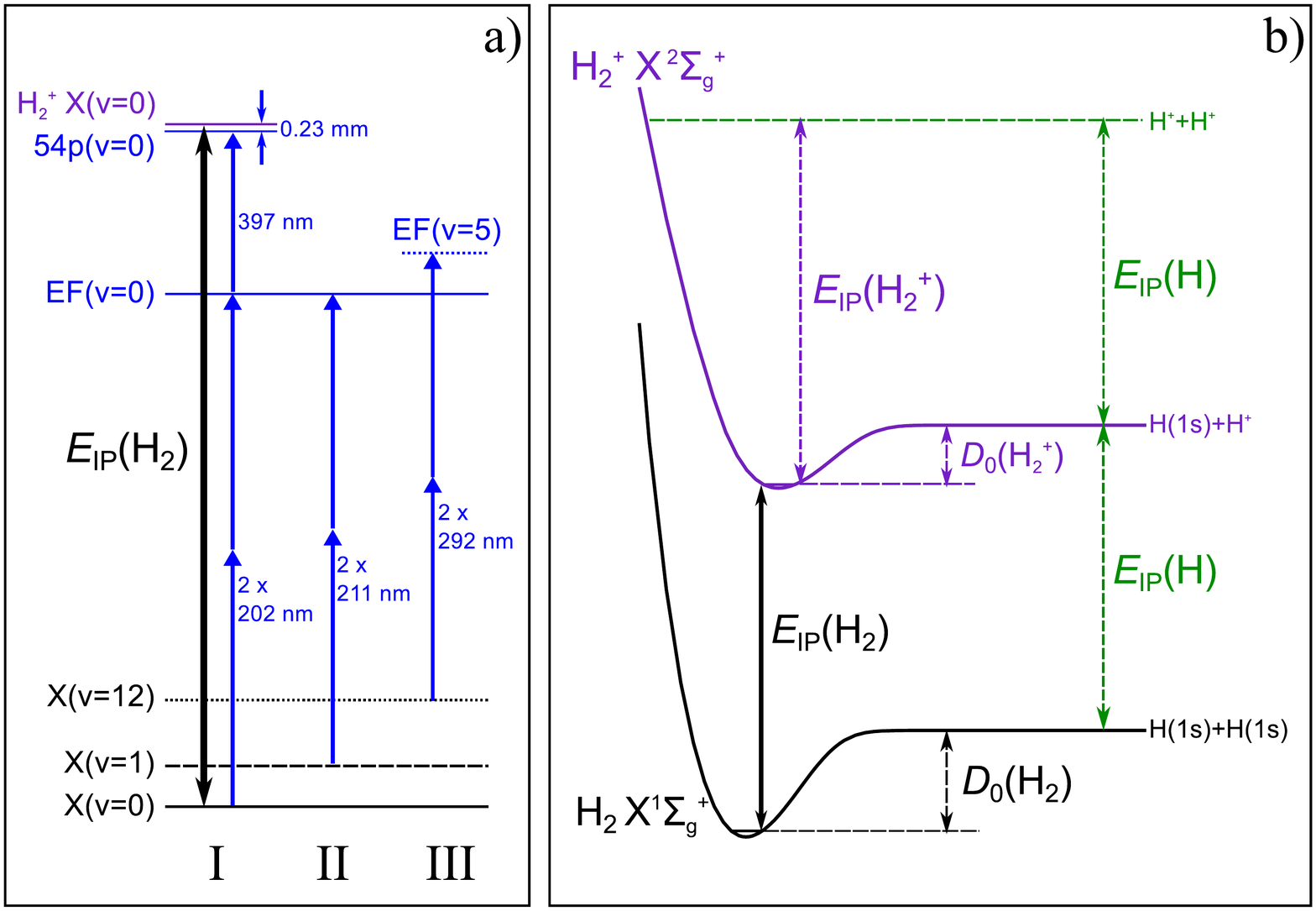}}
\caption{Excitation and level schemes pertaining to the various experiments carried out in neutral hydrogen. a) Three different laser excitation schemes: (I) A three-step excitation method to determine the ionization potential of H$_2$; (II) The two-photon excitation from discharge-populated H$_2$ $v=1$ levels; the combination of I and II delivers values of the fundamental vibrational splitting; (III) Excitation scheme probing the H$_2$ $v=12$ levels populated by photolysis of H$_2$S molecules; b) Potential energy diagram of H$_2$/H$_2^+$ illustrating how a measurement of E$_{\rm{IP}}$(H$_2$) can be converted into a value of D$_0$(H$_2$).
}
\label{H2scheme}
\end{figure}

In recent years great progress is made in both the experimental and theoretical determination of level energies of rovibrational quantum states in the $X^1\Sigma_g^+$ electronic ground state of the hydrogen molecule and its isotopomers. In a combined effort by the Z\"{u}rich and Amsterdam laser precision spectroscopy teams the ionization potential (IP) of the H$_2$ molecule has been measured following scheme I as illustrated in Fig.~\ref{H2scheme}. This work is based on the concept that an accurate value of the dissociation limit can be derived from a measurement of the IP, as was previously shown~\cite{Herzberg1969,Herzberg1972}. The dissociation limit is a well-calculable benchmark property of the canonical molecular quantum system.

Scheme I in Fig.~\ref{H2scheme} represents a combination of three separate measurements to determine the IP. Firstly, measurements were performed on the largest energy splitting between of the \EFX\ system, via deep-UV Doppler-free two-photon laser spectroscopy~\cite{Hannemann2006}.
Secondly, the frequency gap between the $EF^1\Sigma_g^+$ state and a $54p$ Rydberg state was measured in single-photon excitation in a molecular beam at 397 nm~\cite{Liu2009}.
Thirdly, a mm-wave experiment using radiation at wavelengths of 0.23 mm was performed to measure level splittings between Rydberg states and extrapolate to the IP using multi-channel quantum defect theory~\cite{Osterwalder2004}.
The combination of these three independent measurements yields an experimental value for the IP of H$_2$~\cite{Liu2009}. From a measurement of the IP-limit E$_{\rm{IP}}$(H$_2$) a value of the dissociation limit D$_0$(H$_2$) can be derived through a thermodynamic cycle
\begin{equation}
\label{H2-IP-diss}
     \rm{D}_0(\rm{H}_2) = E_{IP}(\rm{H}_2) + D_0(\rm{H}_2^+) - E_{IP}(\rm{H})
\end{equation}
as is also illustrated graphically in Fig.~\ref{H2scheme}b. The dissociation energy of the molecular ion D$_0$(H$_2^+$) can be calculated to high accuracy~\cite{Korobov2008}, while an accurate value for the atomic ionization potential E$_{\rm{IP}}$(H) is known from experiment and theory~\cite{Biraben2009}. By this three-step sequence of precision measurements the ionization potentials and dissociation limits  were determined for the stable isotopomers of neutral hydrogen, H$_2$~\cite{Liu2009}, D$_2$~\cite{Liu2010}, and HD~\cite{Sprecher2010} at an accuracy of $4 \times 10^{-4}$ \wn. Results of these experimental studies are listed in Table~\ref{data}.

An important benchmark for a precision study in neutral molecules is the measurement of the fundamental vibration of the hydrogen molecule, i.e. the energy splitting between $v=0$ and $v=1$ levels in the ground electronic state. Experimental studies thereof had been performed in the past via the very weak quadrupole absorption spectrum in the infrared range~\cite{Buijs1971,Bragg1982} or via the Raman spectrum~\cite{Stoicheff1957,Rahn1990}.
Both approaches suffer from the effect of Doppler broadening, while the infrared measurement is also affected by collisional broadening and shifts, since the weak absorptions can only be observed at elevated pressures.

A novel approach was undertaken to measure the vibrational splitting $v=0 \rightarrow 1$ via the combination difference in electronic transitions \EFX\ for (0,0) and (0,1) bands. The excitation schemes, displayed as I and II in Fig.~\ref{H2scheme}a, sharing a common excited state in the \EF\ ($v=0$) level, were performed with two-photon excitation at $\lambda =202$ nm (from the \X\ $v=0$ ground state) and at $\lambda =210$ nm (from the \X\ $v=1$ ground state, prepared in a discharge pulsed beam source).
The measurements were carried out in the collision-free conditions of a molecular beam, while the Doppler broadening was fully suppressed in a two-beam counter-propagating arrangement~\cite{Hannemann2006}. By this means linewidths of ~36 MHz, much narrower than obtainable in the direct quadrupole transitions, were achieved. Values for the rotationless $J=0$ fundamental vibration for H$_2$, D$_2$ and HD were determined at an accuracy of $2 \times 10^{-4}$ \wn~\cite{Dickenson2013} and listed in Table~\ref{data}.
These splittings were also measured for $J=1$ and $J=2$ for the three isotopomers at the same accuracy~\cite{Niu2014}.

Additional experiments have been performed to determine precise level energies of either rotationally or vibrationally excited quantum states in the $X^1\Sigma_g^+$ ground state potential of H$_2$, again involving two-photon excitation to the \EF\ state. For use in various schemes the rovibrational levels of this first electronically excited \EF\ state, associated with a double-well potential, were calibrated in a preparatory study. As discussed above in section~\ref{muvar}, results from Fourier-transform emission~\cite{Bailly2010,Salumbides2008} were combined with the absolute frequency calibration of the lowest para and ortho levels in the \EF\ state from the two-photon UV measurements as in scheme I~\cite{Hannemann2006}. This combination of methods yields absolute level energies of a large manifold of rovibrational levels of \EF\ with accuracies in some cases approaching $10^{-4}$ \wn. These \EF\ ($v,J$) levels can be used as anchor levels to determine binding energies of \X\ ($v,J$) levels as shown in the following two examples.

\begin{table*}[ht!]
\caption{Results from precision experiments on neutral hydrogen isotopomers and on the deuterated hydrogen ion. $\delta E_{\rm{exp}}$ represents the experimental uncertainty. References are to experimental studies; theoretical values are discussed in main text. $\delta E_{\rm{calc}}$ represents the theoretical uncertainty and $\delta E$ the combined uncertainty taken in quadrature. $\Delta E$ represents the difference between measured and calculated values. All values in MHz.}
\label{data}
\begin{tabular}{l@{\hspace{15pt}}c@{\hspace{15pt}}r@{\hspace{15pt}}cr@{\hspace{15pt}}r@{\hspace{15pt}}r}
\hline
\hline
   species & splitting & $\delta E_{\rm{exp}}$ & Ref. & $\delta E_{\rm{calc}}$  & $\delta E$  &  $\Delta E$ \\
\hline
\Hm	&$v=0, J=6-12$	            &150$^c$  &\cite{Salumbides2011}	&12         &150$^c$      &20	\\
	&$v=0, J=13-16$	            &300$^c$    &\cite{Salumbides2011}	&27         &300$^c$      &90	\\
	&$v=0\rightarrow1$	        &4.5$^a$ &\cite{Dickenson2013}	&2.7	        &5        &7	    \\
	&$v=0\rightarrow2$	        &30	    &\cite{Kassi2014}	    &50             &60    	  &12       \\
	&$v=0\rightarrow3$	        &1.3	&\cite{Cheng2012}	    &75		        &75       &10	    \\
	&$v=0\rightarrow12$	        &105	&\cite{Niu2015b}	    &140            &170      &150$^b$  \\
	&$D_0$			            &12	    &\cite{Liu2009}		    &30		        &3	      &36       \\
\hline
HD	&$v=0\rightarrow1$	        &7$^a$	&\cite{Dickenson2013}	&2.4	        &8	      &4	    \\
	&$D_0$			            &11	    &\cite{Sprecher2010}	&30		        &27	      &32	    \\
\hline
\Dm	&$v=0\rightarrow1$	        &4.5$^a$ &\cite{Dickenson2013}	&2.1            &5        &-0.6	    \\
	&$v=0\rightarrow2$	        &30	    &\cite{Kassi2012}       &12		        &30	      &-12      \\
	&$D_0$			            &21	    &\cite{Liu2010}		    &27		        &30	      &12	    \\
\hline
\Htwoi &$v=0, J=0\rightarrow2$	&2.3	&\cite{Haase2015}	    &0.003		    &2.3	  &-1.0     \\
\hline
\HDi	&$v=0\rightarrow1$	    &0.064	&\cite{Bressel2012}	    &0.002		    &0.064	  &-0.156	\\
	&$v=0\rightarrow4$	        &0.50	&\cite{Koelemeij2007}	&0.008          &0.50     &-0.35	\\
	&$v=0\rightarrow8$	        &0.41	&\cite{Biesheuvel2015}	&0.015		    &0.41     & 0.22    \\
\hline
\hline
\end{tabular}\\[2pt]
{\footnotesize $^a$Value only for $J=0 \rightarrow 0$, larger uncertainties for $J=1 \rightarrow 1$ and $J=2 \rightarrow 2$.}\\
{\footnotesize $^b$Average taken over four measurements for differing rotational states $J=0-3$.}\\
{\footnotesize $^c$Averages taken over the rotational levels involved.}
\end{table*}

One application is the precise measurement of highly excited rotational levels in the $v=0$ ground state manifold of H$_2$ \X\ \cite{Salumbides2011}. The H$_2$ molecular quantum states of high rotational quantum number were prepared via an ultraviolet photolysis reaction~\cite{Aker1989,Kliner1991b}:
\begin{equation}
\label{HBr-phot}
     \mathrm{HBr} + h\nu_{\rm{UV}} \rightarrow \mathrm{H} + \mathrm{Br}
\end{equation}
and a subsequent chemical reaction:
\begin{equation}
\label{HBr-react}
  \mathrm{H} + \mathrm{HBr} \rightarrow \mathrm{H}_2(J^*)  + \mathrm{Br}
\end{equation}
thus giving access to states of up to $J=16$. By making use of the calibrated \EF\ ($v=0, J^*$) anchor levels and exciting these via two-photon Doppler-free spectroscopy, the binding energies of these high rotational states could be experimentally determined at accuracies of 0.005 \wn~\cite{Salumbides2011}.

In the most recent experimental study highly-excited vibrational levels in H$_2$ were investigated lying only some 2000 \wn\ below the dissociation threshold~\cite{Niu2015b}. Production of vibrational states in $v=12$ was accomplished via two-photon UV-photolysis of H$_2$S~\cite{Steadman1989}.  A separate narrowband UV-laser was employed to perform the spectroscopy, displayed as scheme III in Fig.~\ref{H2scheme}. Again two-photon Doppler-free spectroscopy with counter-propagating beams was applied, now exciting the \EF\ $v=5$ levels in H$_2$, corresponding to the third level $F(3)$ in the outer well of the \EF\ double-well potential, for which level energies had been calibrated~\cite{Bailly2010}. This spectroscopic experiment led to values for the binding energies of levels $v=12,J=0-3$ at an accuracy of $3.5 \times 10^{-3}$ \wn~\cite{Niu2015b}.
This accuracy is somewhat lower than for the vibrational ground tone measurements, due to the rather large AC-Stark effects induced in the experiments by the intense laser pulses required when probing the low densities of prepared H$_2$($v^*=12$) states.

The data on the experiments discussed are presented in Table~\ref{data}, which also includes the results of direct measurements of quadrupole transitions in the hydrogen isotopomers, in particular on 
the (2,0) overtone in H$_2$~\cite{Campargue2012,Kassi2014}, the (3,0) overtone in H$_2$~\cite{Hu2012,Cheng2012,Tan2014} and the (2,0) overtone in D$_2$~\cite{Kassi2012}.
For convenience of comparison Table~\ref{data} lists values in units of MHz.

\section{Laser precision metrology of HD$^+$ and H$_2^+$}
\label{H2ionmetrology}

\begin{figure}
\resizebox{1\linewidth}{!}{\includegraphics{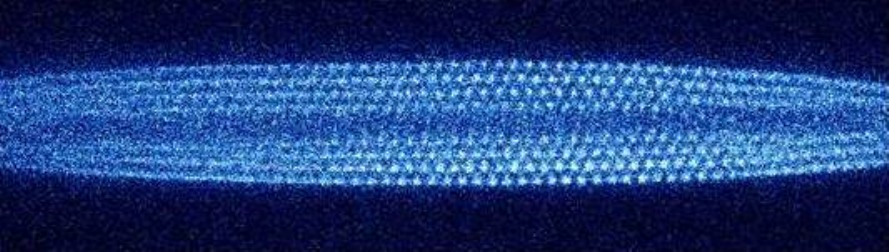}}
\caption{Fluorescence imaging camera picture of a Be$^+$-HD$^+$ two-species sample cooled to below the crystallization temperature forming a Coulomb crystal. Each dot represents a single Be$^+$-ion scattering photons at the 313 nm resonance line used for laser-cooling. The non-fluorescing HD$^+$-ions are confined to the dark band across the centre.}
\label{Cloud}
\end{figure}

The hydrogen molecular ion is an alternative benchmark system for testing first principles QED calculations in experiment~\cite{Hilico2000}.
Being a charged particle it can be subjected to electromagnetic confinement in an ion trap, which has been applied in particular in studies of the HD$^+$ ion~\cite{Roth2006}. Such measurement conditions bear the advantage that extended excitation times can be realized, coping with the weak dipole polarizabilities even in high-order overtone transitions. Moreover, the ions can be sympathetically laser-cooled by adding Be$^+$-ions inside the ion trap, thus reaching mK temperatures at which the Doppler broadening is greatly reduced~\cite{Koelemeij2007}. The laser cooling of two-species ionic ensembles can be pushed to produce Coulomb crystals in which the lighter HD$^+$ particles are confined to the axis of the trap where the electromagnetic fields are zero. The position of the different ions inside a linear Paul trap can be monitored on-line by recording the Be$^+$ fluorescence with an imaging camera as is shown in Fig.~\ref{Cloud}, which was produced with our HD$^+$ ion trap experiment in Amsterdam~\cite{Koelemeij2012}.

\begin{figure}[ht!]
\resizebox{1\linewidth}{!}{\includegraphics{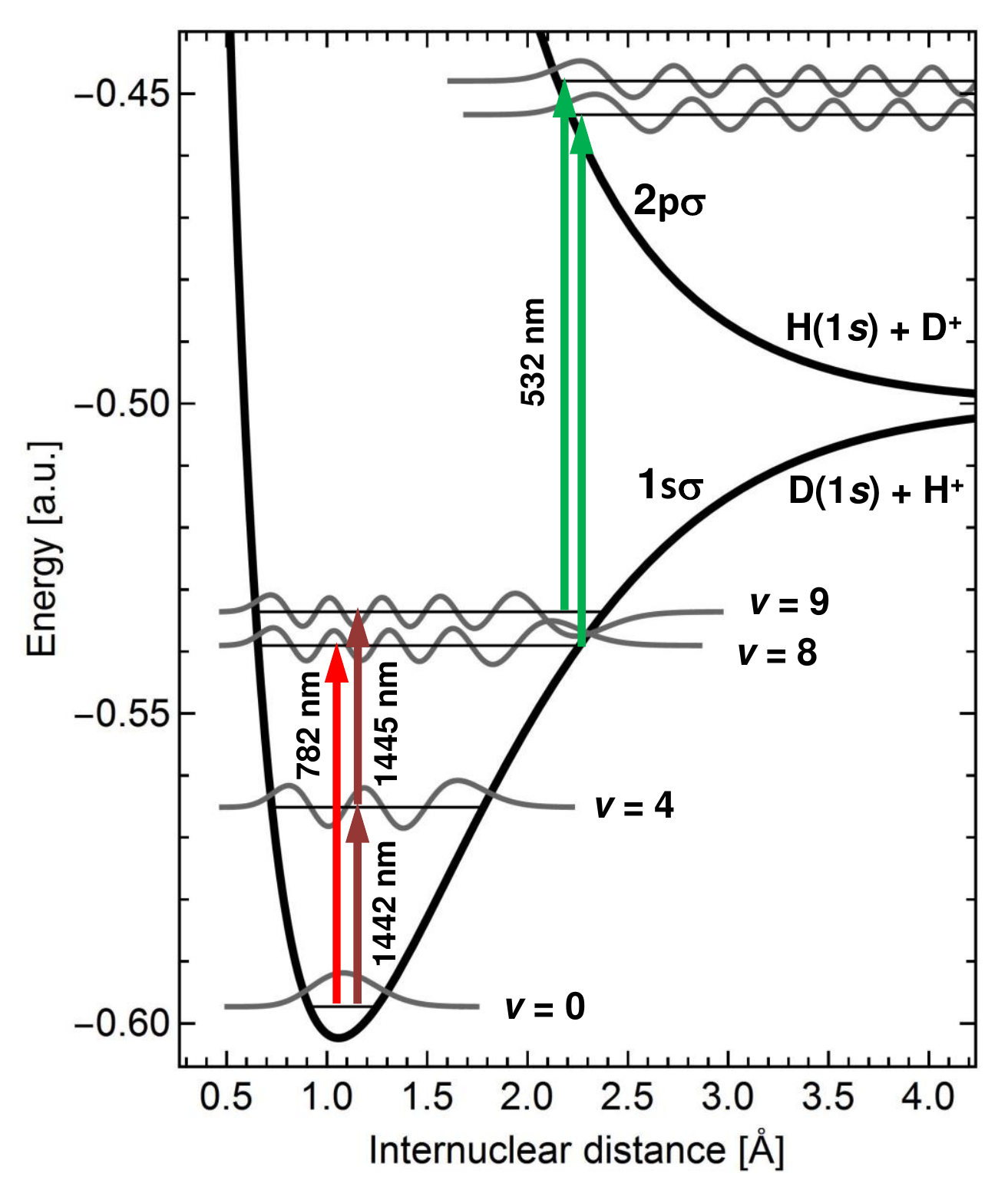}}
\caption{Excitation scheme of resonance-enhanced multi-photon dissociation (REMPD) as employed to detect the spectrum of the R(2) line in the (8,0) band of the HD$^+$ ion at 782 nm. A green laser at 532 nm is used to photolyse the excited HD$^+$ ions via a repulsive state in the ion. Also a proposed two-photon scheme is displayed which will excite the $v=9$ state using the $v=4$ state as a near-resonant intermediate, using lasers at wavelengths of 1442 nm and 1445 nm, to enter the Lamb-Dicke regime to provide Doppler-free conditions.}
\label{HD+}
\end{figure}

The detection of the HD$^+$ resonance is accomplished by measuring the loss of HD$^+$ ions from the trap in a resonance-enhanced multi-photon dissociation (REMPD) scheme as illustrated in Fig~\ref{HD+}. The HD$^+$ ions, excited to the $v=8$ level by a stabilized spectroscopy laser at 782 nm, are dissociated by a second laser at $\lambda=532$ nm~\cite{Koelemeij2012}. For probing the $v=4$ intermediate level the combination 1442 nm/266 nm can be employed~\cite{Roth2006,Koelemeij2007}. The loss of deuterated hydrogen ions can be quantified by a radio-frequent excitation of the Coulomb crystal (secular excitation), therewith liquifying the crystal structure. Under these conditions the intensity of the photons scattered for laser cooling the Be$^+$-ions will increase proportionally to the number of HD$^+$ ions, thus delivering the fluorescence signal to monitor the spectroscopic transition.

Over 50 unresolved hyperfine components contributing to the transition are modeled into a lineshape function, based on the known hyperfine structure for HD$^+$~\cite{Bakalov2006}.
Analyzing a variety of systematic effects, such as the AC and DC Stark effects on the transition frequency~\cite{Koelemeij2011,Karr2014}, the redistribution of rotational sub-states due to black-body radiation in molecular dynamics simulations, and the analysis of hyperfine line strengths at the magnetic sub-state level, the experimental result on the R(2) line of the (8,0) band of HD$^+$ yields a transitions frequency of
$\nu=383,407,177.38\,(41)$ MHz~\cite{Biesheuvel2015}. This result and previous results on transitions in the (4,0)~\cite{Koelemeij2007} and (1,0)~\cite{Bressel2012} bands are included in Table~\ref{data}. Attempts to measure transitions in the (6,0) band have not yet resulted in accurate values~\cite{Zhong2015}.

Work is in progress to measure the spectrum of H$_2^+$ in an ion trap~\cite{Karr2012}, which is more difficult than probing HD$^+$ where electric dipole transitions are allowed, unlike for the homo-nuclear species. Alternatively, the spectroscopy of the H$_2^+$ ion can be approached through measurements on the Rydberg states of the neutral hydrogen molecule, as had been done in the 1990s in measuring the hyperfine structure~\cite{Fu1992}. The Z\"{u}rich group applied laser excitation of Rydberg states of neutral H$_2$ and subsequent millimeter (mm)-wave radiation to probe the level structure of the Rydberg manifolds at principal quantum numbers at $n=50-60$~\cite{Osterwalder2004}. Accurate values for the binding energies of the neutral states could be determined and by application of Multi-channel Quantum Defect Theory an extrapolation could be made for $n \rightarrow \infty$, yielding results on ionic states, for all three isotopomers~\cite{Sprecher2014}.
Remarkably, in these studies the complex hyperfine structure of the ionic cores was better resolved than in the direct measurements on the ions, due to the smaller Doppler broadening achievable in mm-wave spectroscopy.
Recently, this technique was implemented to measure  the $N=0 \rightarrow N=2$ rotational level splitting in para-H$_2^+$~\cite{Haase2015} yielding the accurate value of $5\,223\,485.1\,(2.3)$ MHz included in Table~\ref{data}.

\section{Comparison with QED-calculations}
\label{QED-theory}

Simultaneously with the reports on improved measurements of the dissociation limits of the neutral hydrogen isotopomers H$_2$, HD, and D$_2$, novel \emph{ab initio} calculations on the rovibratonal levels of these species were performed in the framework of non-relativistic quantum electrodynamics and an evaluation in orders of the electromagnetic coupling constant $\alpha$:
\begin{equation}
\label{H2-QED}
    E(\alpha) = E^{(0)} + \alpha^2E^{(2)}  + \alpha^3E^{(3)} + \alpha^4E^{(4)} + \cdots
\end{equation}
The non-relativistic energy $E^{(0)}$ is calculated via an accurate Born-Oppenheimer potential~\cite{Pachucki2010}, where adiabatic and non-adiabatic corrections are calculated in a perturbative approach yielding non-relativistic binding energies for all rovibrational levels~\cite{Pachucki2009}. For the first time accurate relativistic corrections of order $\alpha^2$ and QED-contributions of order $\alpha^3$
were calculated, while an estimate of QED-contributions of order $\alpha^4$ was made, thus leading to an accurate value for the dissociation limits for H$_2$ and D$_2$~\cite{Piszczatowski2009}, and a full set of binding energies of all $(v,J)$ rovibrational levels in the $X^1\Sigma_g^+$ electronic ground state for H$_2$ and D$_2$~\cite{Komasa2011} and for HD~\cite{Pachucki2010b}. Typical accuracies for the binding energies are $10^{-3}$ \wn\ with some variation over the $(v,J)$ set.

\begin{figure}
\resizebox{1\linewidth}{!}{\includegraphics{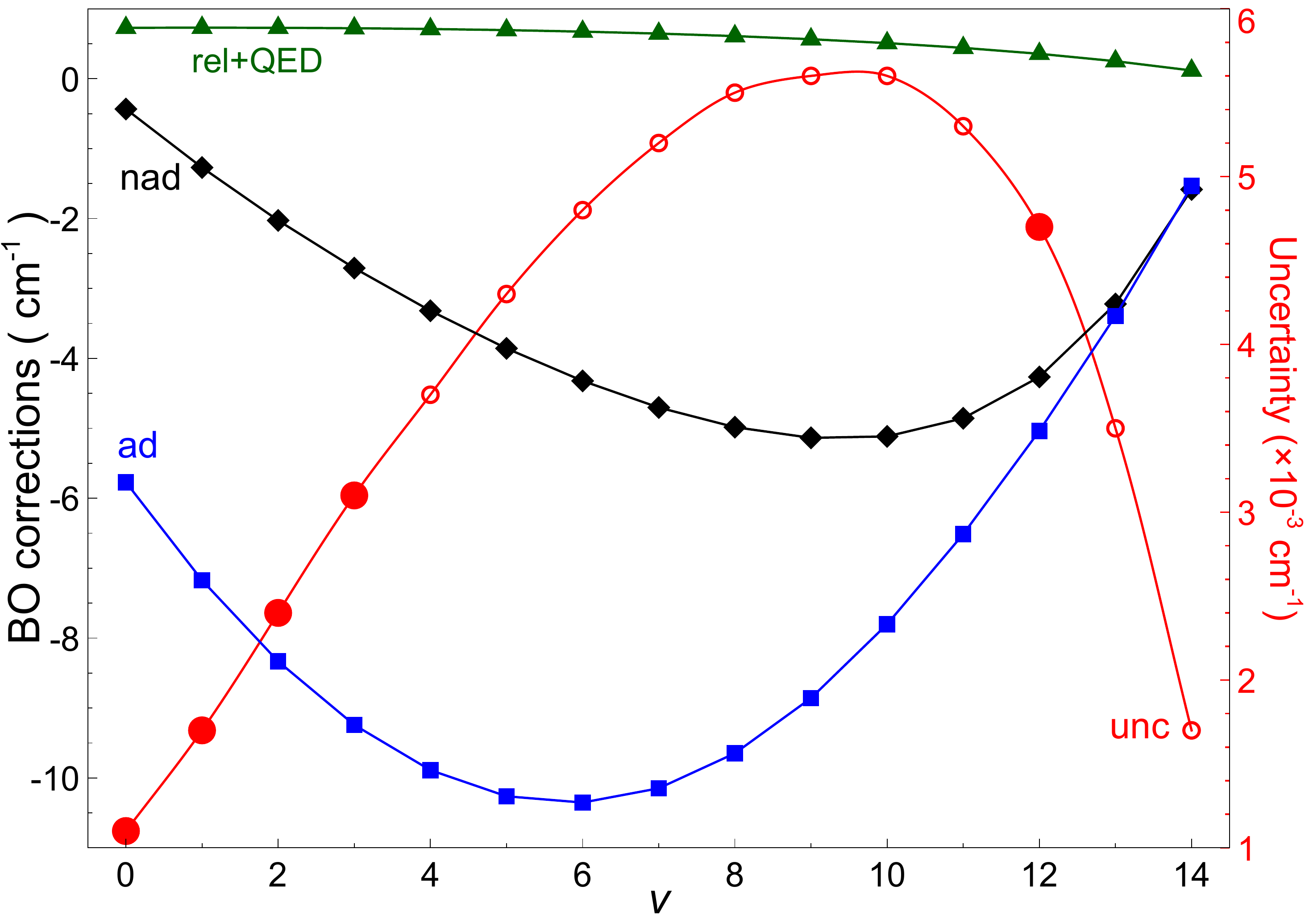}}
\caption{Corrections to the Born-Oppenheimer energies of the H$_2$ \X\ ground electronic state~\cite{Komasa2011} as a function of vibrational quantum number $v$ (in all cases $J=0$). Adiabatic (ad), nonadiabatic (nad), and  relativistic and radiative (rel+QED) corrections are plotted with reference to the scale on the left.
The total uncertainty in the calculations (unc) is plotted with reference to the scale on the right. The thick filled (red) circles relate to measured levels.}
\label{energy_contrib}
\end{figure}

These calculations now stand as the benchmark for comparison with experiments as test of QED in molecular systems.
They constitute an order of magnitude improvement over the previous reference values for the binding energies of ($v,J$) rovibrational levels by Wolniewicz~\cite{Wolniewicz1993,Wolniewicz1995}.
It is for the first time that QED-phenomena in neutral molecules are explicitly calculated, leading to a precision competitive with experiment. In Fig.~\ref{energy_contrib} the values for the correction terms to the Born-Oppenheimer energies, the adiabatic and non-adiabatic as well as the combined relativistic and QED corrections are plotted. Also included in the figure are the uncertainties as currently obtained from first principles calculations. All values in Fig.~\ref{energy_contrib} are restricted to $J=0$ states. The latter values provide insight into approaches to test the QED calculations, with the largest uncertainties remaining for vibrational levels $v=9-10$.

Results from the calculations are also included in Table~\ref{data}. The uncertainties obtained are listed as theoretical uncertainties $\delta E_{\rm{calc}}$ and the deviations between experiment and theory are included as $\Delta E$. Uncertainties from theory and experiment are added in quadrature to obtain values:
\begin{equation}
 \delta E= \sqrt{\delta E_{\rm{calc}}^2 + \delta E_{\rm{exp}}^2}
\end{equation}
which are considered as the 1$\sigma$ combined uncertainties for testing theory. Comparison between the combined uncertainties $\delta E$ and the deviations $\Delta E$ provides a test of QED calculations in these neutral molecular systems. For all cases it is found that $\Delta E < \delta E$, except for the ground tone vibration in H$_2$, where the deviation is just slightly over 1$\sigma$~\cite{Dickenson2013}. For the four measurements on level energies in $v=12$ of H$_2$~\cite{Niu2015b} the deviation $\Delta E$ is averaged and the result is in agreement with QED. This leads us to conclude that QED-theory is found in agreement with the precision measurements on H$_2$, HD and D$_2$, within the uncertainties as listed. This holds for the $J=0$ levels as represented in Table~\ref{data}, but also for the rotational states up to $J=16$~\cite{Salumbides2011}.

It is noted here that the hyperfine structure in the neutral molecules is neglected in this treatment. From the measurements on the hyperfine structure performed by Ramsey and coworkers in the 1950s it was established that the hyperfine splittings for the $J=1$ ortho-levels in H$_2$ is below $0.5$ MHz~\cite{Ramsey1952}, while it is even smaller in the deuterated molecules because of the smaller nuclear $g$-factor of the deuteron, with respect to that of the proton. The para-levels in H$_2$ do not exhibit hyperfine structure. It is therefore safe to assume that hyperfine structure does not play a role for QED-tests in the neutral hydrogen molecules, at the present level of accuracy.
\vspace{0.5cm}

For the HD$^+$ molecular ion the calculations of Moss had long stood as the most accurate~\cite{Moss1993b}. These calculations did not include hyperfine structure and they were performed in the framework of the Born-Oppenheimer approximation.
In recent years, the hyperfine structure of the molecular hydrogen ions was revisited in a series of theoretical works ~\cite{Bakalov2006,Korobov2006,Korobov2009}. The highest precision so far has been achieved for H$_2^+$, and excellent agreement with measured hyperfine intervals~\cite{Jefferts1968,Jefferts1969} was found~\cite{Korobov2015}.

This gives confidence that the hyperfine structure in HD$^+$ is understood as well.
Alternatively, the HD$^+$ ion was considered as a three-body Coulomb system for which variational calculations were performed, bypassing the Born-Oppenheimer approximation~\cite{Korobov2000}. Subsequently leading-order relativistic and radiative corrections were computed~\cite{Korobov2006a}, as well as higher order QED-terms $m\alpha^6$~\cite{Korobov2008} and $m\alpha^7$~\cite{Korobov2014a},
leading to transition frequencies for HD$^+$ below the 0.1 ppb-level~\cite{Korobov2014b}.
This makes these calculations more accurate than those for the neutral hydrogen four-particle molecules where electron correlation plays a complicating role.

Results from calculations for transitions between rovibrational states in HD$^+$ are included in Table~\ref{data} and compared to observations, again deducing $\delta E$ and $\Delta E$ values. While the experimental and theoretical findings for the (4,0)~\cite{Koelemeij2007} and (8,0)~\cite{Biesheuvel2015} bands produce agreement within $1\sigma$, the result of the (1,0) band shows a deviation at the $2.4\sigma$ level. The discrepancy is considered insufficient to call for a disagreement with QED and is treated as an outlier within the full data set of experimental level energies to be compared with theory.

\section{Constraints on physics beyond the Standard Model}
\label{BSM-physics}

In the previous section a number of recent precision measurements on level energies and level splittings in the neutral and ionic hydrogen molecules has led to the conclusion that measurements are in agreement with calculations in the framework of QED. For all entries (except that for the (1,0) transition in HD$^+$) in Table~\ref{data} the relation $\Delta E < \delta E$ was verified. The $\delta E$ values may serve as input for the search for physics beyond the Standard Model (BSM), which can be cast in terms of mathematical tests in which expectation values $\Braket{V_{\rm{BSM}}}$ of quantum mechanical operators relating to new physical phenomena can be calculated and compared with the existing constraints on QED. This concept was recently applied to two possible extensions of the Standard Model, one relating to the possible existence of a fifth force acting between hadronic particles within molecules~\cite{Salumbides2013,Salumbides2014} and one related to the possible existence of extra dimensions~\cite{Salumbides2015b}. Constraints on BSM-physics are then given by the inequality relation:
\begin{equation}
	\Braket{V_{\rm{BSM}}} < \delta E
\label{V5ineq}
\end{equation}

\subsection{Constraints on fifth forces}

In molecular systems gravity is very weak, as the gravitational field between two protons is weaker than the electromagnetic field by $V_G/V_{EM} = 8 \times 10^{-37}$. The strong interaction is confined to femtometer-scale distances and therefore negligible at the \AA\ length scale of a chemical bond in molecules. This nuclear force exhibits an influence on the shape of the electronic wavefunction and hyperfine structure through nuclear finite-size effects. These contributions are well known and readily taken into account. Due to the smallness of these effects, the discrepancy between proton size values obtained from different methods (the proton size puzzle~\cite{Pohl2010}) does not play a significant role in the present work. The weak interaction is also extremely small, below the Hz level, in the hydrogen atom~\cite{Eides2001}, and by proxy it is assumed to be similarly small in the lightest molecules H$_2$ and H$_2^+$. This implies that experimental tests of transition frequencies and level energies as calculated by QED in fact represent tests of the Standard of Model of physics; only the electromagnetic interaction is to be considered when describing the structure of molecules.

The occurrence of a hypothetical fifth force can be expressed in terms of a Yukawa potential of the form~\cite{Salumbides2013}:
\begin{equation}
	V_5(r) = \alpha_5 N_1N_2 \frac{\exp{(-r/\lambda})}{r} \hbar c
\label{V5eq}
\end{equation}
where $\alpha_5$ is the coupling strength of this force, in analogy with the fine structure constant $\alpha_{EM}$, and $\lambda$ is the range of the force. In a quantum field approach this force is then mediated by the exchange of force-carrying bosonic particles of mass $m_5=\hbar/\lambda c$. In the following $\hbar=c=1$ is adopted and we restrict the treatment to hypothetical fifth forces between hadronic particles for which nucleon numbers $N_1$ and $N_2$ are included, $N=1$ for the proton and $N=2$ for the deuteron.

The effect of the fifth force on the energy of a molecular state $\Psi_{v,J}$ can be calculated as an expectation value
\begin{equation}
	\Braket{ V_{5,\lambda} } =  \alpha_5 N_1 N_2  \Braket{ Y_{\lambda} }
\label{V5expect}
\end{equation}
with
\begin{equation}
   \Braket{ Y_{\lambda}} = \Braket{ \Psi_{v,J}(r)  | \frac{\exp{(-r/\lambda})}{r}  | \Psi_{v,J}(r) }
\end{equation}
In this treatment the range of the force $\lambda$ is retained as a parameter, while the expectation value can be numerically evaluated by inserting the well-known wave functions $\Psi_{v,J}(r)$ for the quantum states of the H$_2$ molecules, their isotopomers, and the HD$^+$ ion, by integrating over the coordinate $r$ representing the internuclear distance in the molecule.
This effect of a fifth force on level energies can be extended to transition frequencies by calculating the differences in expectation values for ground states $\Psi_{v'',J''}$ and excited states $\Psi_{v',J'}$, represented by $\Delta Y_{\lambda}$.

\begin{figure}
\resizebox{1\linewidth}{!}{\includegraphics{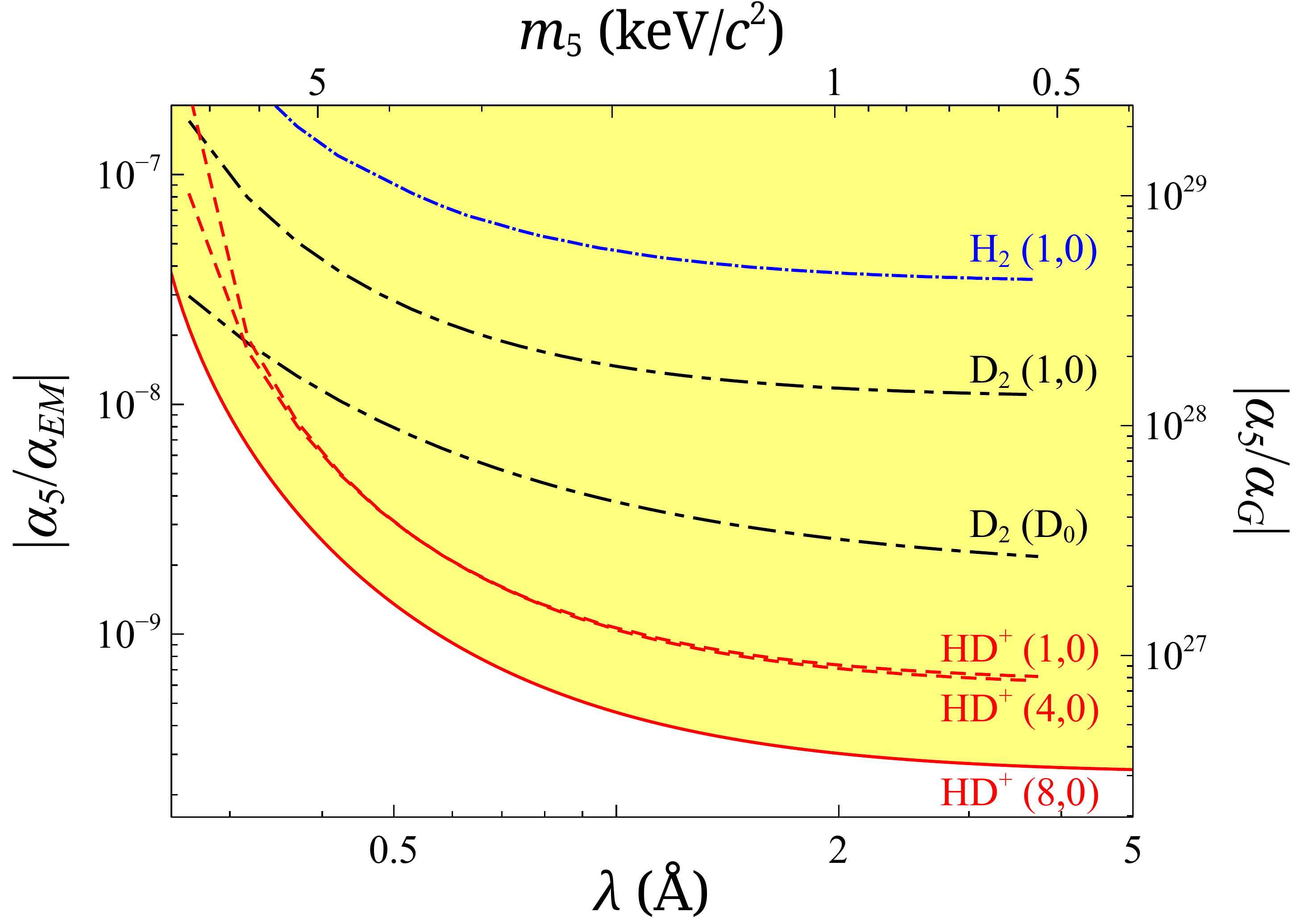}}
\caption{Limits on the strength $\alpha_5$ of a putative fifth force relative to $\alpha_{EM}$ (left scale) or $\alpha_G$ (right scale) as a function of the range parameter $\lambda$ (lower axis). The shaded (yellow) area is the region that is excluded by the molecular precision experiments, with the result obtained in the (8,0) band of HD$^+$.
The top axis indicates the mass $m_5$ of a bosonic carrier of the fifth force.}
\label{limit-5th}
\end{figure}

Constraints on a fifth force can be derived by combining Eq.~(\ref{V5expect}) and Eq.~(\ref{V5ineq}) to~\cite{Salumbides2015b}:
\begin{equation}
   \alpha_5 < \frac{\delta E }{N_1N_2\Delta Y_{\lambda}}
\end{equation}
and  computing the fifth force expectation value for each specific precision measurement and for each value of the range parameter $\lambda$. This then results in the constraints as graphically represented in Fig.~\ref{limit-5th}. The figure shows that the result of the measurement on the (8,0) high overtone band HD$^+$ yields the strongest constraint. This is caused by the fact that the wave function densities and average internuclear distance of $v=8$ deviate more strongly from the ground state wave function $v=0$ in comparison to the other wave functions probed, $v=1$~\cite{Bressel2012} and $v=4$~\cite{Koelemeij2007}. The measurements in HD$^+$ are currently more accurate than for H$_2$, but the effect of the smaller equilibrium distance in H$_2$ with respect to HD$^+$ is reflected in a steeper decline of the constraint for the ions below 0.3 \AA. Note also that constraints from D$_2$ are tighter than those from H$_2$ because of the $N_1N_2$ factor. Note, however, that this is based on the assumption that protons and neutrons deliver an additive contribution to the fifth force acting between hadrons.
The figure also shows how the values for the dissociation energy D$_0$ are more constraining than those of the fundamental vibration measured at higher accuracy. This is because the dissociation limit corresponds to a state with $r \rightarrow \infty$.

\subsection{Constraints on extra dimensions}

A similar reasoning can be followed when considering the effect of extra dimensions.  A theory was developed by Arkani-Hamed and coworkers~\cite{ArkaniHamed1998,Antoniadis1998}, known as ADD-theory, postulating that the three forces of the Standard Model act in three dimensions (on a 3-brane), while gravity may act in a higher ($3+n_e$)-dimensional space, where $n_e$ refers to the number of extra dimensions. As it turns out the effects of gravity become larger in higher dimensional space.
The extra dimensions, if existent, should not extend over large distances otherwise it would be in contradiction with observations.
As was postulated already in 1926 by Klein the extra dimensions might exist nevertheless in compactified form~\cite{Klein1926}.
In our analysis we assume further that the entire molecule of size $R_{mol} \sim 1$ \AA\ is contained within the compactification length $R_c$ of the extra dimensions, so that it effectively probes the higher-dimensional space. This assumption can be made, since the bounds on extra dimensions have not yet been experimentally limited to this molecular scale. Indeed, as the outcome of the present analysis we find  $R_c > R_{mol}$.

Salumbides~\emph{et al.}~\cite{Salumbides2015b} applied the concepts of ADD-theory in a phenomenological fashion to diatomic molecules where the two nuclei are considered to act as test bodies in a Cavendish-type gravity experiment. Again, as in the case of the fifth force, the expectation value $\Braket{V_{ADD}}$ of this potential is calculated in a perturbative manner by integrating over the molecular wave function $\Psi(r)$. While in the previous analysis~\cite{Salumbides2015b} geometrical prefactors, related to volumes and surfaces of hyperspheres, had been ignored, these prefactors will be included here, giving rise to a small numerical adaptation of results.

The flux conservation law for the gravitational force $\vec{F}$ or the gravitational field $\vec{g}=\vec{F}/m$ acting on a test mass $m$ can be recast in $n$ dimensions as
\begin{equation}
   \oint_{\mathcal{V}_n} \frac{\vec{F}}{m} \cdot d\vec{\mathcal{A}_{n}} = - \hat{\mathcal{A}_{n}} G_{n} M
\label{flux-law}
\end{equation}
where $G_{n}$ is the gravitational constant and $M$ the mass contained within the spherical volume $\mathcal{V}_{n}$.
Assuming that $\vec{F}$ is spherically symmetric (equivalent to the requirement that $\vec{\nabla}\times\vec{F}=0$), the choice of an $n$-dimensional hypersphere as the enclosing Gaussian volume simplifies the integration.
The integral is taken over the hypersurface surrounding the $n$-dimensional hypersphere, defined as an $(n-1)$ dimensional surface $\mathcal{A}_{n}$, which relates to the volume  $\mathcal{V}_n$ of the $n$-dimensional hypersphere with radius $r$ via $\mathcal{A}_{n}=n\mathcal{V}_n/r$, yielding:
\begin{equation}
  \mathcal{A}_{n} = \hat{\mathcal{A}_{n}} r^{n-1} = \left[ \frac{n\pi^{n/2}}{\Gamma(\frac{n}{2}+1)} \right]  r^{n-1},
\label{hypersurface}
\end{equation}
where $\hat{\mathcal{A}}_{n}$ is the surface area for a unit hypersphere ($r=1$).

This then yields an equation for Newton's law in $n$ dimensions as:
\begin{equation}
  F_n = -G_{n} \frac{Mm}{r^{(n-1)}},
\label{Newtons-law}
\end{equation}
where the Newtonian inverse square law dependence is recovered for 3 dimensions.
Note that $G_{n}$ have different dimensional units when compared to the (3-dimensional) gravitational constant $G_{3}=G$.
Following the analysis of \cite{Arkani-Hamed1999}, assuming that the $n_e$ extra spatial dimensions have the same compactification length $R_c$, the $n$-dimensional force $F_n$ can be derived in the limiting case when $r \gg R_c$ to be
\begin{equation}
  F_n = -\left\{ \frac{n\pi^{n/2}}{4\pi \Gamma(\frac{n}{2}+1)}    \frac{G_{n} }{(R_{c})^{n_e} } \right\} \frac{Mm}{r^{2}}.
\label{Newtons-law-correspondence}
\end{equation}
This should be in correspondence with normal Newtonian gravity, i.e. $F_{n} \rightarrow F_{3}$ as $R_{c}/r \rightarrow 0$, which implies that the factor inside the braces $\{...\}$ is equivalent to the 3-dimensional gravitational constant $G$.
Indeed the definition for the gravitational constant $(G_n)$ for the case of $n$-dimensions:
\begin{equation}
  G_n = \left[ \frac{4\pi \Gamma(\frac{n}{2}+1)}{n\pi^{n/2}} \right] { (R_{c})^{n_e} } G
\label{Gn-factor}
\end{equation}
is the very solution invoked in ADD theory~\cite{Arkani-Hamed1999} to solve the hierarchy problem on the weakness of gravity compared to the other SM interactions.

The $n$-dimensional gravitational potential can be derived via the relation $\vec{F_n}= - {\nabla} V_n$, yielding
\begin{equation}
  V_n = - \left[ \frac{4\pi \Gamma(\frac{n}{2}+1)}{n(n-2)\pi^{n/2}} \right] { (R_{c})^{n_e} } G  \frac{M m }{r^{(n-2)}},
\label{n-pot}
\end{equation}
where $G_n$ has been expressed in terms of $G$.
Note that the geometrical prefactor [$f$] for the potential:
\begin{equation}
 [f_n] =  \left[ \frac{4\pi \Gamma(\frac{n}{2}+1)}{n(n-2)\pi^{n/2}} \right]
\label{prefactor}
\end{equation}
scales differently as a function of $n$, but reduces to [$f_3$]=1 for 3-dimensional space. However for extra dimensions the prefactor becomes [$f_4$]=$1/\pi$, [$f_5$]=$1/2\pi$, [$f_6$]=$1/\pi^2$, etc.

The extra-dimensional gravitational potential of ADD-theory (within the compactification radius $r<R_c$) is expressed in a convenient form for application in molecular theory as
\begin{equation}
   V_{ADD}(r)= -\alpha_G N_1N_2 R_c^{n_e} [f_{(n_e + 3)}] \frac{1}{r^{(n_e+1)}},
\end{equation}
where $M=m=m_p$ is set to be the nucleon mass(es), $\alpha_{G} = G m_p^2/\hbar c=5.9\times10^{-39}$ represents the very small gravitational coupling constant, and $N_1$ and $N_2$ the number of hadrons (protons and neutrons) in the attracting bodies.
Outside the compactification radius ($r>R_c$) gravity is assumed to follow the 3-dimensional Newtonian law:
\begin{equation}
   V_N(r)= -G\frac{Mm}{r} = -\alpha_G N_1N_2 \frac{1}{r}
\end{equation}
which will also be used for scaling purposes.

The perturbative effect on a quantum state is then, again following~\cite{Salumbides2015b}:
\begin{eqnarray}
   \Braket{V_{ADD}(n_e,R_c)} &=& -\alpha_G N_1N_2 R_c^{n_e} [f_{(n_e+3)}] \times  \nonumber \\
	& &\Braket{\Psi(r) | \frac{1}{r^{(n_e+1)}} | \Psi(r) }
\end{eqnarray}
where the number of extra dimensions $n_e$ and the compactification radius $R_c$ (taken equal for all extra dimensions) are retained as parameters.
Similarly as in the case of the fifth force treatment a perturbative effect can also be calculated on a transition frequency by taking the differing perturbations on the ground and excited quantum states.

\begin{figure}
\resizebox{1\linewidth}{!}{\includegraphics{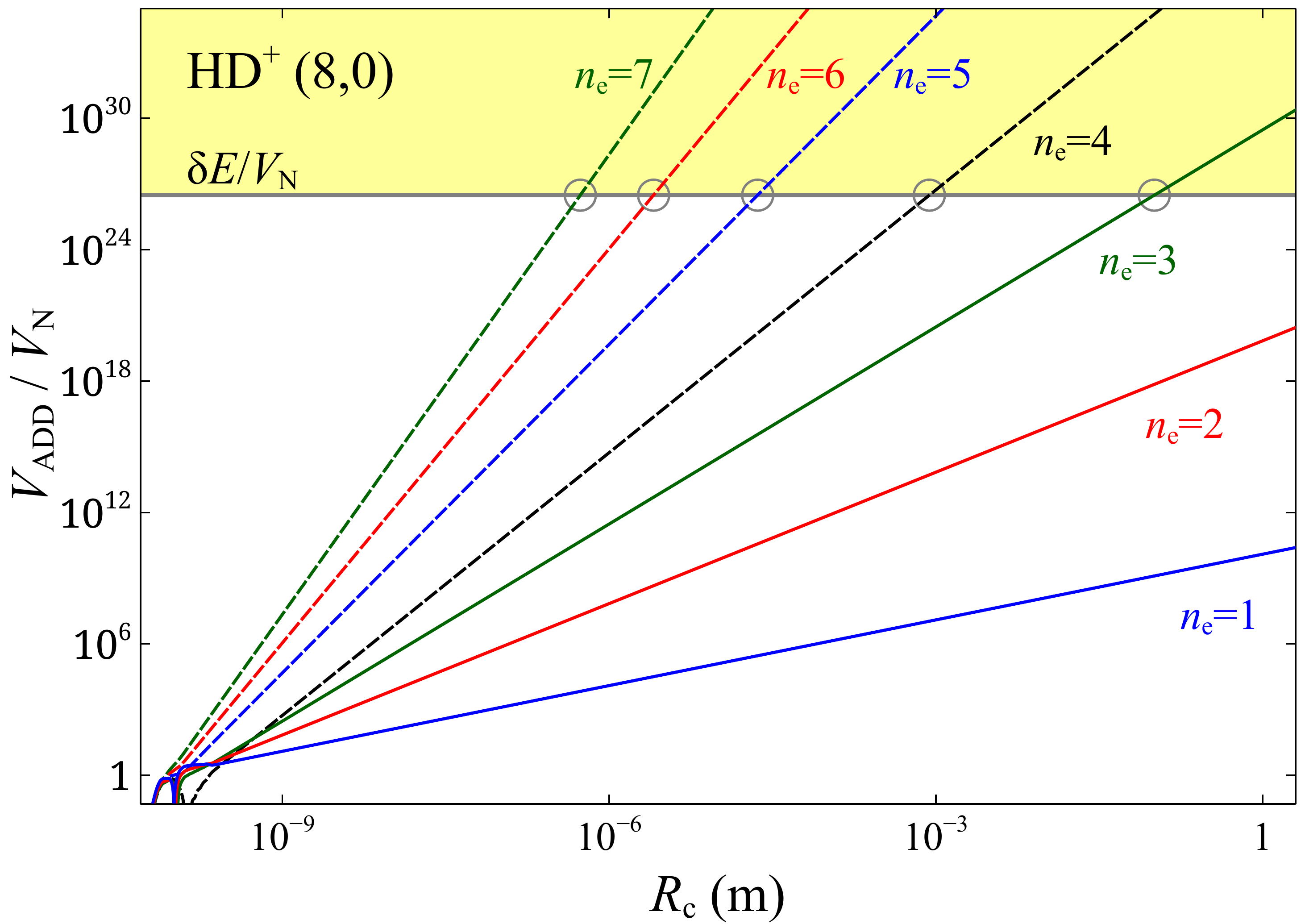}}
\caption{Upper bounds on the compactification radius as obtained from the most constraining molecular spectroscopy result, i.e. the  observed transition in the HD$^+$ (8,0) band. The shaded (yellow) region is excluded based on the combined uncertainty from measurement and theory $\delta E$. The straight lines represent the expectation values $\Braket{V_{ADD}}$ for different numbers of extra dimensions $n_e$. Both $\delta E$ and $\Braket{V_{ADD}}$ are scaled to the Newtonian potential $V_N$. The crossing points yield the bounds of compactification radii, measuring $R_c < 0.6$ $\mu$m for the case of 7 extra dimensions of relevance to M-theory.}
\label{limit-dim}
\end{figure}

The experimental data are then implemented, as prescribed in Eq.~(\ref{V5ineq}), to constrain the sizes of the extra dimensions for each number of extra dimensions $n_e$ by:
\begin{eqnarray}
   (R_c)^{n_e} &<& \frac{\delta E}{\alpha_G N_1N_2 [f_{(n_e+3)}]} \times  \nonumber \\
	& &{\left[ \Braket{\frac{1}{r^{n_e+1}}}_{\Psi_e} - \Braket{\frac{1}{r^{n_e+1}}}_{\Psi_g} \right]^{-1}}
\end{eqnarray}
with calculation of the expectation values for ground $\Psi_g$ and excited $\Psi_e$ states in a transition.
Hence constraints on the compactification radius $R_c$ can be deduced for cases of differing $n_e$ extra dimensions. For the special case of the measurement of the (8,0) band of HD$^+$ results are graphically displayed in Fig.~\ref{limit-dim}. In the figure the extra-dimensional contribution to gravity as represented by $\Braket{V_{ADD}}$ is scaled to normal gravity $V_N$, thus demonstrating that the effect of the former is much larger.

From Fig.~\ref{limit-dim} a constraint on a compactification radius can be read for each number $n_e$. In particular for the special case of the M-theory which lives in 10 spatial dimensions~\cite{Witten1995} a constraint follows of $R_c<0.6\,\mu$m, under the assumption that all extra dimensions have equal extension.

\section{Outlook and perspective}
\label{outlook}

Some 10 H$_2$ absorption systems at high redshift in the range $z=2.0-4.2$ have been analyzed from spectra observed with the best large dish optical telescopes of the 8-10m class (ESO Very Large Telescope at Paranal, Chile, and the Keck telescope at Hawaii) equipped with high resolution spectrometers resulting in a constraint on a varying proton-electron mass ratio of $|\Delta\mu/\mu|< 5 \times 10^{-6}$ at 3$\sigma$. With current technologies no decisive improvements are expected, although in a quest for varying constants even incremental improvements are valuable. A number of 23 additional H$_2$ absorbers have been identified~\cite{Ubachs2015}, some of which apparently of the same quality as the 10 already analyzed in detail. Another set of 23 systems is reported exhibiting certain or only tentative detection of H$_2$~\cite{Balashev2014}. Improvements on the quality of data can be accomplished with the planned larger dish telescopes (the European Extremely Large Telescope, the Thirty-Meter-Telescope, and the Giant Magellan Telescope) and with the use of fiber-fed spectrometers (like ESPRESSO~\cite{Pepe2010} planned for the VLT), as well as the use of frequency-comb assisted frequency calibration~\cite{Murphy2007a,Wilken2012}. This should lead to at least an order-of-magnitude tighter constraint on a varying $\mu$.

Detection of H$_2^+$ ions in space, either at high redshift or galactic, might impose similar constraints on varying constants. Attempts to detect these ionic forms of hydrogen molecules have  been unsuccessful so far, despite dedicated searches by radio astronomy on the spin-splitting resonance at $\sim1.4$ GHz~\cite{Penzias1968,Varshalovich1993a}. Also searches for vibrationally excited states have been performed~\cite{Shuter1986}, in view of the fact that the H$_2^+$ to H$_3^+$ conversion is slower for vibrationally excited states.
The development of new radio observatories of extreme sensitivity as the Square Kilometer Array~\cite{Dewdney2009} bears the prospect of detecting H$_2^+$ in outer space.

It is remarkable that none of the electronically excited states of the H$_2^+$ ion, calculated accurately already in the early days of quantum mechanics by Teller~\cite{Teller1930} has ever been spectroscopically observed, neither in space nor in the laboratory. This is due to the fact that the electronically excited states ($3d\sigma$, $2p\pi$, $3d\pi$, etc.) have potential wells lying at such large internuclear separation that only Franck-Condon allowed decay is possible to very highest vibrational levels and the dissociation continuum of the $1s\sigma$ ground state. Such decay would produce radiation of $\sim 10$ eV, hence in the vacuum ultraviolet part of the spectrum. An exception is the $2p\sigma$ state which exhibits a strongly repulsive well, but supports nevertheless some weakly bound states at larger internuclear separation, which were observed through microwave spectroscopy in H$_2^+$~\cite{Carrington1989} and in HD$^+$~\cite{Carrington1988b}.

\begin{figure}
\resizebox{1\linewidth}{!}{\includegraphics{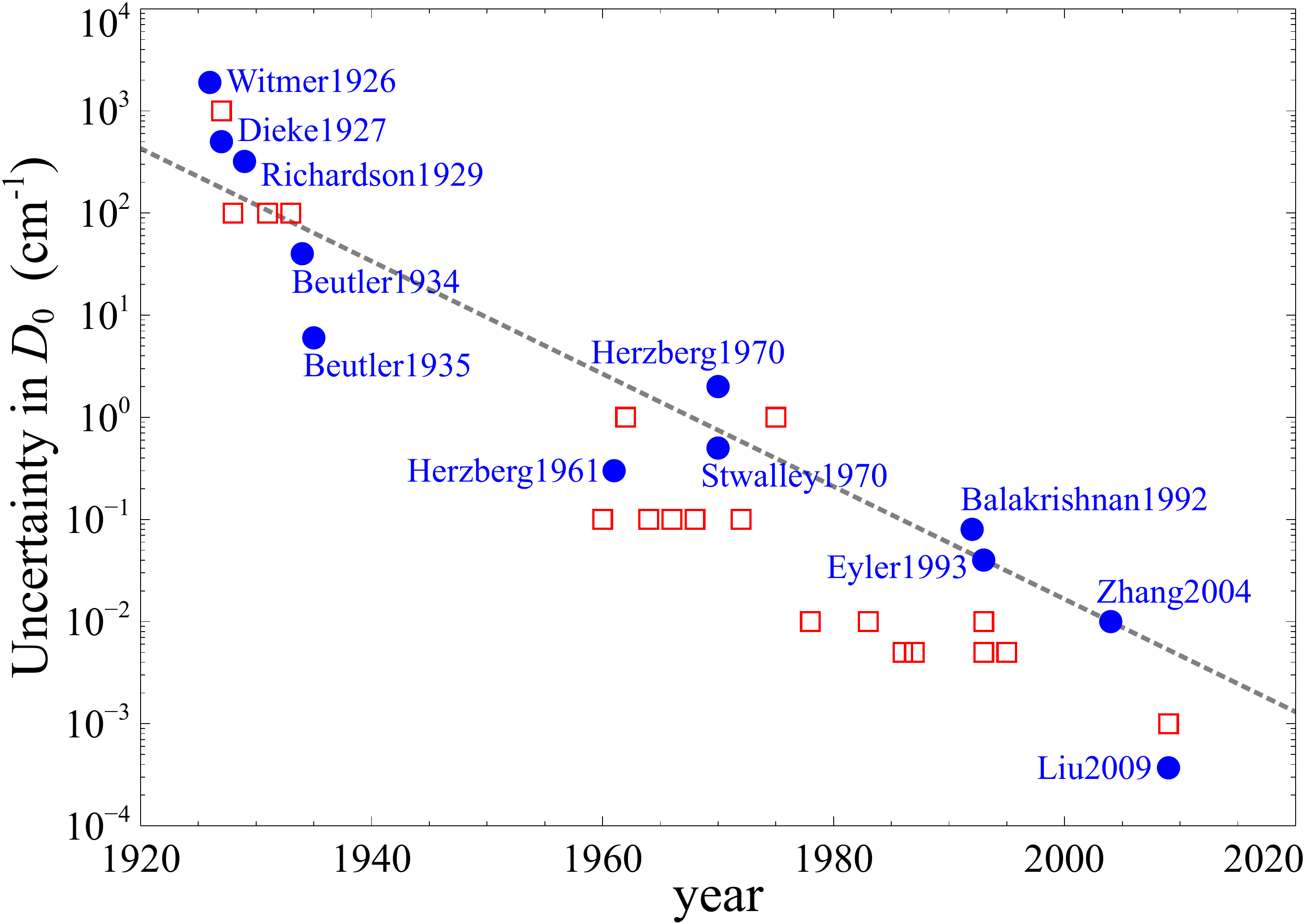}}
\caption{Historical progress on measurements (filled (blue) circles) and theoretical calculations (open (red) squares) of the dissociation energy of the hydrogen molecule. Plotted are the uncertainties in D$_0$, rather than the values.
The theoretical data are, in chronological order, from references~\cite{Heitler1927,Sugiura1927,Wang1928,Rosen1931,James1933,Kolos1960,Kolos1962,Kolos1964,Wolniewicz1966,Kolos1968,Bunker1972,Kolos1975,Bishop1978,Wolniewicz1983,Kolos1986,Schwartz1987,Kolos1993,Wolniewicz1993,Wolniewicz1995,Piszczatowski2009},
not explicitly identified.
The experimental data are identified and are associated with references~\cite{Witmer1926,Dieke1927,Richardson1929,Beutler1934,Beutler1935,Herzberg1961,Herzberg1970,Stwalley1970,Balakrishnan1992,Eyler1993,Zhang2004,Liu2009}.
Data collection based on Sprecher {et al.}~\cite{Sprecher2011}.}
\label{Fig-dis}
\end{figure}

As for laboratory measurements it is noted that, since the advent of quantum mechanics, progress in measurements and theory of the dissociation limit of H$_2$ has been mutually stimulating, therewith producing steady progress. This history, both for theoretical and experimental advances, is presented in Fig.~\ref{Fig-dis} in terms of the accuracy of the values determined. The data on theory starts with the calculation by Heitler and London~\cite{Heitler1927}
mainly aiming at the quantum mechanical explanation that the ground state of the H$_2$ is bound. Thereafter a great number of calculations were performed of ever increasing accuracy. As is common for ab initio calculations, uncertainties were not explicitly stated, and in Fig.~\ref{Fig-dis} estimates of uncertainties were derived from the number of digits given in the original papers. However, in view of the fact that QED-effects were not explicitly calculated the accuracies are estimated to be limited to $5 \times 10^{-3}$ \wn~\cite{Kolos1986,Schwartz1987,Kolos1993,Wolniewicz1995}. The results of some of those calculations appeared to be accurate, but that is due to the fact that not-addressed relativistic and QED-contributions cancel for a large part.
Only for the final insert the ab initio calculation explicitly includes relativistic and QED-effects up to $\alpha^4$~\cite{Piszczatowski2009} achieving an accuracy of $1.0 \times 10^{-3}$ \wn.

The first experimental determination of the hydrogen dissociation energy was established just before the advent of quantum mechanical calculations in 1926~\cite{Witmer1926}. This measurement was superseded over the years by experimental studies of increasing accuracy
with the most accurate measurement achieving a precision of $4 \times 10^{-4}$ \wn~\cite{Liu2009}.
Fig.~\ref{Fig-dis} demonstrates how the improvement in the accuracy of this benchmark value of molecular physics has increased by some seven orders of magnitude in less than a century, and how the improvement of experiment and theory have gone hand-in-hand.

For the future a number of possible improvements may be identified.
Through the measurement of the binding energy of the $GK^1\Sigma_g^+, v=1$ state~\cite{Sprecher2013} a novel route for bridging the gap between the electronic ground state and the IP-limit is opened, as an alternative to the \EF, $v=0$ state as an intermediate (see also Fig.~\ref{H2scheme}a). This novel route involves only a two-step process, instead of the three-step process via \EF. It requires however Doppler-free 2-photon excitation at $\lambda=179$ nm, which could possibly be generated at sufficient intensity by harmonic upconversion in a KBBF crystal~\cite{Zhang2008}.
Another prospect for improvement is in the remeasurement of the anchor lines, either of the \EF\ state or the \GK\ state, using Ramsey-comb spectroscopy with ultrafast laser pulses \cite{Morgenweg2014,Morgenweg2014b}. In this case excitation takes place with two ultrafast phase-locked laser pulses derived from a frequency comb laser. Ramsey spectroscopy signal can be obtained by varying the delay between the pulses on two different time scales via adjustments of the frequency comb laser. From a series of Ramsey signals the transition frequencies can be obtained with possibly two orders of magnitude improved accuracy ($<100$ kHz) compared to the single-nanosecond pulse excitation technique used so far.

On the theory side improvements have been made recently in accurate calculations of H$_2$ level energies, in particular for the adiabatic corrections~\cite{Pachucki2014} and the non-adiabatic corrections~\cite{Pachucki2015}. For a comparison with experiment and more stringent test of QED in molecular systems further improvements on the calculations of relativistic and QED contributions are awaited.
Experimental work on H$_2$ at increasing accuracy may also provide a handle to target $u-g$ symmetry-breaking effects in the homo-nuclear species, calculated to be at the $10^{-6}$ \wn\ level~\cite{Pachucki2011} for the very highest vibrational states, which might be probed by the photolysis preparation methods explored~\cite{Niu2015b}.
Tests of QED in neutral molecules can be extended to tritium containing nuclei, already included in the calculations, while only few experimental studies have been performed~\cite{Veirs1987,Chuang1987}.

For the more distant future it may be noted that the lifetimes of the ($v,J$) rovibrational levels of the $X^1\Sigma_g^+$ electronic ground state of H$_2$ are in the range of $10^6$ seconds~\cite{Black1976,Pachucki2011a}. When performing measurements at infrared frequencies, i.e. probing quadrupole transitions, of some $10^{14}$ Hz, the spectral lines exhibit a natural line width of $10^{-20}$ of the transition frequency, which might in principle allow for a 20-digit determination of level energies to be compared with theory. Of course many experimental difficulties need to be overcome to reach such accuracies, where activities in decelerating, cooling and trapping of H$_2$~\cite{Hogan2009,Seiler2011} may become useful ingredients.
In any case there exists a vast "room-at-the-bottom" for improving precision metrology of neutral hydrogen in search for new physics.

For the charged hydrogen molecule the precision achieved in theory is much higher, with calculations up to order $\alpha^7$, while experiments on HD$^+$ are currently the most accurate, and the most constraining for testing new physics (see e.g. Fig.~\ref{limit-5th}).
Prospects are that accuracy can be improved and proposals for molecular clocks based on hydrogen molecular ions are discussed~\cite{Karr2014,Schiller2014}.
A value of the proton-electron mass ratio $\mu$ can ultimately be determined from spectroscopy, as was already postulated long time ago by Wing \emph{et al.}~\cite{Wing1976}. Calculations of level energies have recently reached a precision at which they are limited by the value of $\mu$ rather than the theoretical methods, while the sensitivity on other fundamental constants such as the deuteron-proton mass ratio and the proton size has become significant~\cite{Korobov2014a,Korobov2014b}. Experiments for the charged species are however much less accurate than theory, even with the trapping and laser-cooling methods applied.

Spectroscopy of improved precision in the HD$^+$-ion trap can be achieved by Doppler-free two-photon laser excitation with counter-propagating beams. Transition probabilities were found to be weak using degenerate two-photon excitation~\cite{Karr2005}.
A scheme based on two-photon excitation with near-degenerate laser beams was proposed bearing the advantage of increased transition probabilities~\cite{Tran2013}.
In a scheme where two different wavelengths are imposed (see Fig.~\ref{HD+}), involving a first the transition $v=0 \rightarrow v=4$ at $\lambda=1442$ nm and a second transition  $v=4 \rightarrow v=9$ at $\lambda=1445$ nm, the Lamb-Dicke regime can be reached. This makes this  measurement measurement scheme essentially  Doppler-free if counter-propagating laser beams are used. Calculations and numerical simulations have shown that accuracies at the $10^{-14}$ level can be reached~\cite{Tran2013}, therewith providing a greatly enhanced accuracy on the transition frequencies in HD$^+$. This will produce much tighter constraints on fifth forces and extra dimensions, in addition to a competitive value of the proton-electron mass ratio $\mu$.

\section*{Acknowledgements}
The authors wish to thank Julija Bagdonaite, John Barrow, Martin Barstow, Denise Bailly, Jurriaan Biesheuvel, Mario Dapr\`{a}, Gareth Dickenson, Beatriz Gato-Rivera, Laurent Hilico, Urs Hollenstein, Toncho Ivanov, Christian Jungen, Lex Kaper, Jean-Philippe Karr, Julian King, Jacek Komasa, Vladimir Korobov, Jinjun Liu, Frederic Merkt, Michael Murphy, Mingli Niu, Krzystof Pachucki, Simon Preval, Mourad Roudjane, Evelyne Roueff, Bert Schellekens, Daniel Sprecher, Lydia Tchang-Brillet, Michel Vervloet and John Webb for their collaboration and the fruitful discussions.
This work is financially supported by the Netherlands Foundation for Fundamental Research of Matter (FOM) through the program "Broken Mirrors \& Drifting Constants" and by the European Research Council (ERC) under the European
Union's Horizon 2020 research and innovation programme (grant agreement No 670168).


\begin{thebibliography}{166}
\expandafter\ifx\csname url\endcsname\relax
  \def\url#1{\texttt{#1}}\fi
\expandafter\ifx\csname urlprefix\endcsname\relax\def\urlprefix{URL }\fi
\expandafter\ifx\csname href\endcsname\relax
  \def\href#1#2{#2} \def\path#1{#1}\fi

\bibitem{Atlas2012}
{Atlas collaboration}, Observation of a new particle in the search for the
  {S}tandard {M}odel {H}iggs boson with the {ATLAS} detector at the {LHC},
  Phys. Lett. B 716 (2012) 1 -- 29.

\bibitem{CMS2012}
{CMS collaboration}, Observation of a new boson at a mass of 125 {GeV} with the
  {CMS} experiment at the {LHC}, Phys. Lett. B 716 (2012) 30 -- 61.

\bibitem{Hudson2011}
J.~J. Hudson, D.~M. Kara, I.~J. Smallman, B.~E. Sauer, M.~R. Tarbutt, E.~A.
  Hinds, Improved measurement of the shape of the electron, Nature 473 (2011)
  493--496.

\bibitem{ACME2014}
{ACME collaboration}, Order of magnitude smaller limit on the electric dipole
  moment of the electron, Science 343 (2014) 269--272.

\bibitem{Loh2013}
H.~Loh, K.~C. Cossel, M.~C. Grau, K.~K. Ni, E.~R. Meyer, J.~L. Bohn, J.~Ye,
  E.~A. Cornell, {Precision spectroscopy of polarized molecules in an ion
  trap}, Science 342 (2013) 1220--1222.

\bibitem{Deangelis1996}
M.~de~Angelis, G.~Gagliardi, L.~Gianfrani, G.~M. Tino, Test of the
  symmetrization postulate for spin-0 particles, Phys. Rev. Lett. 76 (1996)
  2840--2843.

\bibitem{Naus1997}
H.~Naus, A.~de~Lange, W.~Ubachs,
  ${b}^{1}{\ensuremath{\Sigma}}_{g}^{+}-{X}^{3}{\ensuremath{\Sigma}}_{g}^{-}$
  (0,0) band of oxygen isotopomers in relation to tests of the symmetrization
  postulate in ${}^{16}${O}$_{2}$, Phys. Rev. A 56 (1997) 4755--4763.

\bibitem{Gianfrani1999}
L.~Gianfrani, R.~W. Fox, L.~Hollberg, Cavity-enhanced absorption spectroscopy
  of molecular oxygen, J. Opt. Soc. Am. B 16 (1999) 2247--2254.

\bibitem{Modugno1998}
G.~Modugno, M.~Inguscio, G.~M. Tino, Search for small violations of the
  symmetrization postulate for spin-0 particles, Phys. Rev. Lett. 81 (1998)
  4790--4793.

\bibitem{Mazotti2001}
D.~Mazzotti, P.~Cancio, G.~Giusfredi, M.~Inguscio, P.~De~Natale, Search for
  exchange-antisymmetric states for spin-0 particles at the ${10}^{-11}$ level,
  Phys. Rev. Lett. 86 (2001) 1919--1922.

\bibitem{Bekenstein2002}
J.~D. {Bekenstein}, {Fine-structure constant variability, equivalence
  principle, and cosmology}, \prd 66 (2002) 123514.

\bibitem{Barrow2002a}
J.~Barrow, H.~Sandvik, J.~Magueijo, Behavior of varying-$\alpha$ cosmologies,
  \prd 65 (2002) 063504.

\bibitem{Carroll1998}
S.~M. {Carroll}, {Quintessence and the rest of the world: suppressing
  long-range interactions}, \prl 81 (1998) 3067--3070.

\bibitem{Bagdonaite2015}
J.~Bagdonaite, W.~Ubachs, M.~T. Murphy, J.~B. Whitmore, Constraint on a varying
  proton-electron mass ratio 1.5 billion years after the big bang, \prl 114
  (2015) 071301.

\bibitem{Lamb1947}
W.~E. Lamb, R.~C. Retherford, Fine structure of the hydrogen atom by a
  microwave method, \pr 72 (1947) 241--243.

\bibitem{Jansen2014}
P.~{Jansen}, H.~L. {Bethlem}, W.~{Ubachs}, {Perspective: Tipping the scales:
  Search for drifting constants from molecular spectra}, \jcp 140 (2014)
  010901.

\bibitem{Ubachs2007}
W.~{Ubachs}, R.~{Buning}, K.~S.~E. {Eikema}, E.~{Reinhold}, {On a possible
  variation of the proton-to-electron mass ratio: H$_{2}$ spectra in the line
  of sight of high-redshift quasars and in the laboratory}, \jms 241 (2007)
  155--179.

\bibitem{Philip2003}
J.~Philip, J.~Sprengers, T.~Pielage, C.~A. de~Lange, W.~Ubachs, E.~Reinhold,
  Highly accurate transition frequencies in the {H}$_{2}$ {L}yman and {W}erner
  absorption bands, Can. J. Chem. 82 (2004) 713--722.

\bibitem{Ubachs2004}
W.~Ubachs, E.~Reinhold, Highly accurate {H$_2$} {L}yman and {W}erner band
  laboratory measurements and an improved constraint on a cosmological
  variation of the proton-to-electron mass ratio, \prl 92 (2004) 101302.

\bibitem{Reinhold2006}
E.~{Reinhold}, R.~{Buning}, U.~{Hollenstein}, A.~{Ivanchik}, P.~{Petitjean},
  W.~{Ubachs}, {Indication of a cosmological variation of the proton-electron
  mass ratio based on laboratory measurement and reanalysis of H$_{2}$
  spectra}, \prl 96 (2006) 151101.

\bibitem{Hollenstein2006}
U.~{Hollenstein}, E.~{Reinhold}, C.~A. {de Lange}, W.~{Ubachs},
  {High-resolution {VUV}-laser spectroscopic study of the
  {B$^1\Sigma^+_{(u)}(v' = 0-2) \leftarrow X^1\Sigma^+_{(g)}(v'' = 0)$} {L}yman
  bands in {H$_2$} and {HD}}, \jpb 39 (2006) L195--L201.

\bibitem{Hannemann2006}
S.~Hannemann, E.~J. Salumbides, S.~Witte, R.~T. Zinkstok, E.~J. van Duijn,
  K.~S.~E. Eikema, W.~Ubachs, Frequency metrology on the ${EF}^{1}{\Sigma}
  _{g}^{+}-{X}^{1}{\Sigma}_{g}^{+}$ (0,0) transition in {H}$_{2}$, {HD}, and
  {D}$_{2}$, Phys. Rev. A 74 (2006) 062514.

\bibitem{Bailly2010}
D.~Bailly, E.~Salumbides, M.~Vervloet, W.~Ubachs, Accurate level energies in
  the {${EF}^{1}{\Sigma}^{+}_{g}$}, {${GK}^{1}{\Sigma}^{+}_{g}$},
  {${H}^{1}{\Sigma}^{+}_{g}$}, ${B}^{1}{\Sigma}^{+}_{u}$, ${C}^{1}{\Pi}_{u}$,
  ${B'}^{1}{\Sigma}^{+}_{u}$, ${D}^{1}{\Pi}_{u}$, ${I}^{1}{\Pi}_{g}$,
  ${J}^{1}{\Delta}_{g}$ states of {H}$_{2}$, \molp 108 (2010) 827--846.

\bibitem{Salumbides2008}
E.~J. {Salumbides}, D.~{Bailly}, A.~{Khramov}, A.~L. {Wolf}, K.~S.~E. {Eikema},
  M.~{Vervloet}, W.~{Ubachs}, {Improved laboratory values of the H$_{2}$ Lyman
  and Werner lines for constraining time variation of the proton-to-electron
  mass ratio}, \prl 101 (2008) 223001.

\bibitem{Ivanov2008a}
T.~I. {Ivanov}, M.~{Roudjane}, M.~O. {Vieitez}, C.~A. {de Lange}, W.-{\"U}.~L.
  {Tchang-Brillet}, W.~{Ubachs}, {HD as a probe for detecting mass variation on
  a cosmological time scale}, \prl 100 (2008) 093007.

\bibitem{Ivanov2010}
T.~I. {Ivanov}, G.~D. {Dickenson}, M.~{Roudjane}, N.~D. {Oliveira},
  D.~{Joyeux}, L.~{Nahon}, W.-U.~L. {Tchang-Brillet}, W.~{Ubachs},
  Fourier-transform spectroscopy of {HD} in the vacuum ultraviolet at
  {$\lambda$} = 87-112 nm, \molp 108 (2010) 771--786.

\bibitem{Malec2010}
A.~L. {Malec}, R.~{Buning}, M.~T. {Murphy}, N.~{Milutinovic}, S.~L. {Ellison},
  J.~X. {Prochaska}, L.~{Kaper}, J.~{Tumlinson}, R.~F. {Carswell}, W.~{Ubachs},
  {Keck telescope constraint on cosmological variation of the
  proton-to-electron mass ratio}, \mnras 403 (2010) 1541--1555.

\bibitem{Weerdenburg2011}
F.~{van Weerdenburg}, M.~T. {Murphy}, A.~L. {Malec}, L.~{Kaper}, W.~{Ubachs},
  {First constraint on cosmological variation of the proton-to-electron mass
  ratio from two independent telescopes}, \prl 106 (2011) 180802.

\bibitem{Bagdonaite2012}
J.~{Bagdonaite}, M.~T. {Murphy}, L.~{Kaper}, W.~{Ubachs}, {Constraint on a
  variation of the proton-to-electron mass ratio from H$_{2}$ absorption
  towards quasar Q2348$-$011}, \mnras 421 (2012) 419--425.

\bibitem{Bagdonaite2014a}
J.~{Bagdonaite}, W.~{Ubachs}, M.~T. {Murphy}, J.~B. {Whitmore}, {Analysis of
  molecular hydrogen absorption toward QSO B0642$-$5038 for a varying
  proton-to-electron mass ratio}, \apj 782 (2014) 10.

\bibitem{Dapra2015}
M.~{Dapr\`{a}}, J.~{Bagdonaite}, M.~T. {Murphy}, W.~{Ubachs}, {Constraint on a
  varying proton-to-electron mass ratio from molecular hydrogen absorption
  toward quasar SDSS J123714.60+064759.5}, \mnras 454 (2015) 489--506.

\bibitem{King2011}
J.~A. {King}, M.~T. {Murphy}, W.~{Ubachs}, J.~K. {Webb}, {New constraint on
  cosmological variation of the proton-to-electron mass ratio from
  Q0528$-$250}, \mnras 417 (2011) 3010--3024.

\bibitem{Rahmani2013}
H.~{Rahmani}, M.~{Wendt}, R.~{Srianand}, P.~{Noterdaeme}, P.~{Petitjean},
  P.~{Molaro}, J.~B. {Whitmore}, M.~T. {Murphy}, M.~{Centurion},
  H.~{Fathivavsari}, S.~{D'Odorico}, T.~M. {Evans}, S.~A. {Levshakov},
  S.~{Lopez}, C.~J.~A.~P. {Martins}, D.~{Reimers}, G.~{Vladilo}, {The UVES
  large program for testing fundamental physics - II. Constraints on a change
  in {$\mu$} towards quasar HE 0027$-$1836}, \mnras 435 (2013) 861--878.

\bibitem{King2008}
J.~A. {King}, J.~K. {Webb}, M.~T. {Murphy}, R.~F. {Carswell}, {Stringent null
  constraint on cosmological evolution of the proton-to-electron mass ratio},
  \prl 101 (2008) 251304.

\bibitem{Thompson2009}
R.~I. {Thompson}, J.~{Bechtold}, J.~H. {Black}, D.~{Eisenstein}, X.~{Fan},
  R.~C. {Kennicutt}, C.~{Martins}, J.~X. {Prochaska}, Y.~L. {Shirley}, {An
  observational determination of the proton-to-electron mass ratio in the early
  universe}, \apj 703 (2009) 1648--1662.

\bibitem{Wendt2012}
M.~{Wendt}, P.~{Molaro}, {QSO 0347$-$383 and the invariance of m$_{p}$/m$_{e}$
  in the course of cosmic time}, \aap 541 (2012) A69.

\bibitem{Varshalovich2001}
D.~A. {Varshalovich}, A.~V. {Ivanchik}, P.~{Petitjean}, R.~{Srianand},
  C.~{Ledoux}, {HD molecular lines in an absorption system at redshift $z =
  2.3377$}, Astron. Lett. 27 (2001) 683--685.

\bibitem{Khoury2004}
J.~{Khoury}, A.~{Weltman}, {Chameleon fields: Awaiting surprises for tests of
  gravity in space}, \prl 93 (2004) 171104.

\bibitem{Xu2013}
S.~{Xu}, M.~{Jura}, D.~{Koester}, B.~{Klein}, B.~{Zuckerman}, {Discovery of
  molecular hydrogen in white dwarf atmospheres}, \apjl 766 (2013) L18.

\bibitem{Bagdonaite2014b}
J.~{Bagdonaite}, E.~J. {Salumbides}, S.~P. {Preval}, M.~A. {Barstow}, J.~D.
  {Barrow}, M.~T. {Murphy}, W.~{Ubachs}, {Limits on a gravitational field
  dependence of the proton-electron mass ratio from H$_{2}$ in white dwarf
  stars}, \prl 113 (2014) 123002.

\bibitem{Salumbides2015}
E.~J. Salumbides, J.~Bagdonaite, H.~Abgrall, E.~Roueff, W.~Ubachs, {H$_2$}
  {L}yman and {W}erner band lines and their sensitivity for a variation of the
  protonâ€“electron mass ratio in the gravitational potential of white dwarfs,
  \mnras 450 (2015) 1237--1245.

\bibitem{Herzberg1969}
G.~Herzberg, Dissociation energy and ionization potential of molecular
  hydrogen, Phys. Rev. Lett. 23 (1969) 1081--1083.

\bibitem{Herzberg1972}
G.~Herzberg, C.~Jungen, Rydberg series and ionization potential of the {H$_2$}
  molecule, J. Mol. Spectrosc. 41 (1972) 425--486.

\bibitem{Liu2009}
J.~Liu, E.~J. Salumbides, U.~Hollenstein, J.~C.~J. Koelemeij, K.~S.~E. Eikema,
  W.~Ubachs, F.~Merkt, Determination of the ionization and dissociation
  energies of the hydrogen molecule, \jcp 130 (2009) 174306.

\bibitem{Osterwalder2004}
A.~Osterwalder, A.~{W\"{u}est}, F.~Merkt, C.~Jungen, High-resolution millimeter
  wave spectroscopy and multichannel quantum defect theory of the hyperfine
  structure in high {R}ydberg states of molecular hydrogen {H$_2$}, \jcp 121
  (2004) 11810--11838.

\bibitem{Korobov2008}
V.~I. Korobov, Relativistic corrections of {$m\alpha^6$} order to the
  rovibrational spectrum of {H$_2^+$} and {HD$^+$} molecular ions, Phys. Rev. A
  77 (2008) 022509.

\bibitem{Biraben2009}
F.~Biraben, Spectroscopy of atomic hydrogen. {H}ow is the {R}ydberg constant
  determined, Eur. Phys. J. Special Topics 172 (2009) 109--119.

\bibitem{Liu2010}
J.~Liu, D.~Sprecher, C.~Jungen, W.~Ubachs, F.~Merkt, Determination of the
  ionization and dissociation energies of the deuterium molecule ({D}$_{2}$),
  \jcp 132 (2010) 154301.

\bibitem{Sprecher2010}
D.~Sprecher, J.~Liu, C.~Jungen, W.~Ubachs, F.~Merkt, Communication: The
  ionization and dissociation energies of {HD}, \jcp 133 (2010) 111102.

\bibitem{Buijs1971}
H.~Buijs, G.~Bush, Static field induced spectrum of hydrogen, \cjp 49 (1971)
  2366--2375.

\bibitem{Bragg1982}
S.~L. Bragg, W.~H. Smith, J.~W. Brault, Line positions and strengths in the
  {H$_2$} quadrupole spectrum, \apj 263 (1982) 999--1004.

\bibitem{Stoicheff1957}
B.~P. Stoicheff, High resolution {R}aman spectroscopy of gases {IX.} {H$_{2}$,
  HD and D$_{2}$}, Can. J. Phys. 35 (1957) 730--741.

\bibitem{Rahn1990}
L.~Rahn, G.~Rosasco, Measurement of the density shift of the {H}$_{2}$ {Q}(0-5)
  transitions from 295 to 1000 {K}, Phys. Rev. A 41 (1990) 3698--3706.

\bibitem{Dickenson2013}
G.~D. Dickenson, M.~L. Niu, E.~J. Salumbides, J.~Komasa, K.~S.~E. Eikema,
  K.~Pachucki, W.~Ubachs, Fundamental vibration of molecular hydrogen, \prl 110
  (2013) 193601.

\bibitem{Niu2014}
M.~L. Niu, E.~J. Salumbides, G.~D. Dickenson, K.~S.~E. Eikema, W.~Ubachs,
  Precision spectroscopy of the {$X^1\Sigma_g^+, v=0 \rightarrow 1\, (J=0-2$)}
  rovibrational splittings in {H$_2$}, {HD} and {D$_2$}, \jms 300 (2014)
  44--54.

\bibitem{Salumbides2011}
E.~J. Salumbides, G.~D. Dickenson, T.~I. Ivanov, W.~Ubachs, {QED} effects in
  molecules: Test on rotational quantum states of {H}$_{2}$, \prl 107 (2011)
  043005.

\bibitem{Kassi2014}
S.~Kassi, A.~Campargue, Electric quadrupole transitions and collision-induced
  absorption in the region of the first overtone band of {H$_2$} near 1.25
  {$\mu$m}, \jms 300 (2014) 55 -- 59.

\bibitem{Cheng2012}
C.-F. Cheng, Y.~R. Sun, H.~Pan, J.~Wang, A.-W. Liu, A.~Campargue, S.-M. Hu,
  Electric-quadrupole transition of {H$_{2}$} determined to {10$^{-9}$}
  precision, \pra 85 (2012) 024501.

\bibitem{Niu2015b}
M.~L. Niu, E.~J. Salumbides, W.~Ubachs, Communication: Test of quantum
  chemistry in vibrationally hot hydrogen molecules, \jcp 143 (2015) 081102.

\bibitem{Kassi2012}
S.~Kassi, A.~Campargue, K.~Pachucki, J.~Komasa, The absorption spectrum of
  {D$_2$}: {U}ltrasensitive cavity ring down spectroscopy of the (2–0) band
  near 1.7 {$\mu$}m and accurate ab initio line list up to 24 000 cm{$^{-1}$},
  \jcp 136 (2012) 184309.

\bibitem{Haase2015}
C.~Haase, M.~Beyer, C.~Jungen, F.~Merkt, The fundamental rotational interval of
  para-{H$_2^+$} by {MQDT}-assisted {R}ydberg spectroscopy of {H$_2$}, \jcp 142
  (2015) 064310.

\bibitem{Bressel2012}
U.~Bressel, A.~Borodin, J.~Shen, M.~Hansen, I.~Ernsting, S.~Schiller,
  Manipulation of individual hyperfine states in cold trapped molecular ions
  and application to {HD$^{+}$} frequency metrology, Phys. Rev. Lett. 108
  (2012) 183003.

\bibitem{Koelemeij2007}
J.~C.~J. Koelemeij, B.~Roth, A.~Wicht, I.~Ernsting, S.~Schiller, Vibrational
  spectroscopy of {HD$^{+}$} with 2-ppb accuracy, Phys. Rev. Lett. 98 (2007)
  173002.

\bibitem{Biesheuvel2015}
J.~Biesheuvel, J.-P. Karr, L.~Hilico, K.~S.~E. Eikema, W.~Ubachs, J.~C.~J.
  Koelemeij, Molecular vibrations as a probe of fundamental physical constants
  and laws, Submitted.

\bibitem{Aker1989}
P.~M. Aker, G.~J. Germann, J.~J. Valentini, State-to-state dynamics of {H}+{HX}
  collisions. {I}. the {H}+{HX} $\rightarrow$ {H}$_{2}$+{X} ({X=Cl,Br,I})
  abstraction reactions at 1.6 e{V} collision energy, \jcp 90 (1989)
  4795--4808.

\bibitem{Kliner1991b}
D.~A.~V. Kliner, D.~E. Adelman, R.~N. Zare, Comparison of experimental and
  theoretical integral cross sections for {D+H$_2$}{($v=1, J=1$)} -
  {HD}{($v’=1, J’$)}+{H}, \jcp 95 (1991) 1648--1662.

\bibitem{Steadman1989}
J.~Steadman, T.~Baer, The production and characterization by resonance enhanced
  multiphoton ionization of {H$_2$} {($v=10-14$)} from photodissociation of
  {H$_2$S}, \jcp 91 (1989) 6113--6119.

\bibitem{Campargue2012}
A.~Campargue, S.~Kassi, K.~Pachucki, J.~Komasa, The absorption spectrum of
  {H}$_{2}$: {CRDS} measurements of the (2-0) band, review of the literature
  data and accurate ab initio line list up to 35 000 cm$^{-1}$, Phys. Chem.
  Chem. Phys. 14 (2012) 802--815.

\bibitem{Hu2012}
S.-M. Hu, H.~Pan, C.-F. Cheng, Y.~R. Sun, X.-F. Li, J.~Wang, A.~Campargue,
  A.-W. Liu, The {$v = 3 - 0$} {S}(0)-{S}(3) electric quadrupole transitions of
  {H$_{2}$} near 0.8 {$\mu$m}, Astroph. J. 749 (2012) 76.

\bibitem{Tan2014}
Y.~Tan, J.~Wang, C.-F. Cheng, X.-Q. Zhao, A.-W. Liu, S.-M. Hu, Cavity ring-down
  spectroscopy of the electric quadrupole transitions of in the 784 {$–-$} 852
  nm region, \jms 300 (2014) 60--64.

\bibitem{Hilico2000}
L.~Hilico, N.~Billy, B.~{Gr\'{e}maud}, D.~Delande, Ab initio calculation of the
  {$J = 0$} and {$J = 1$} states of the {H$_2^+$}, {D$_2^+$} and {HD$^+$}
  molecular ions, \eujpd 12 (2000) 449--466.

\bibitem{Roth2006}
B.~Roth, J.~C.~J. Koelemeij, H.~Daerr, S.~Schiller, Rovibrational spectroscopy
  of trapped molecular hydrogen ions at millikelvin temperatures, Phys. Rev. A
  74 (2006) 040501.

\bibitem{Koelemeij2012}
J.~C.~J. Koelemeij, D.~W.~E. Noom, D.~de~Jong, M.~A. Haddad, W.~Ubachs,
  Observation of the {$vâ€²=8 \leftarrow v=0$} vibrational overtone in cold
  trapped {HD$^+$}, \apb 107 (2012) 1075--1085.

\bibitem{Bakalov2006}
D.~Bakalov, V.~I. Korobov, S.~Schiller, High-precision calculation of the
  hyperfine structure of the {HD$^{+}$} ion, Phys. Rev. Lett. 97 (2006) 243001.

\bibitem{Koelemeij2011}
J.~C.~J. Koelemeij, Infrared dynamic polarizability of {HD$^+$} rovibrational
  states, \pccp 13 (2011) 18844--18851.

\bibitem{Karr2014}
J.-P. Karr, {H$_2^+$} and {HD$^+$}: {C}andidates for a molecular clock, \jms
  300 (2014) 37--43.

\bibitem{Zhong2015}
Z.-X. Zhong, X.~Tong, Z.-C. Yan, T.-Y. Shi, High-precision spectroscopy of
  hydrogen molecular ions, Chin. Phys. B 24 (2015) 053102.

\bibitem{Karr2012}
J.-P. Karr, A.~Douillet, L.~Hilico, Photodissociation of trapped {H$_2^+$} ions
  for {REMPD} spectroscopy, \apb 107 (2012) 1043--1052.

\bibitem{Fu1992}
Z.~W. Fu, E.~A. Hessels, S.~R. Lundeen, Determination of the hyperfine
  structure of {H$_2^+$} ({$\nu=0, R =1$}) by microwave spectroscopy of
  high-{$L, n =27$} {R}ydberg states of {H$_2$}, Phys. Rev. A 46 (1992)
  R5313--R5316.

\bibitem{Sprecher2014}
D.~Sprecher, C.~Jungen, F.~Merkt, Determination of the binding energies of the
  {$np$} {R}ydberg states of {H$_2$}, {HD}, and {D$_2$} from high-resolution
  spectroscopic data by {M}ultichannel {Q}uantum-{D}efect {T}heory, \jcp 140
  (2014) 104303.

\bibitem{Pachucki2010}
K.~Pachucki, Born-{O}ppenheimer potential for {H$_2$}, Phys. Rev. A 82 (2010)
  032509.

\bibitem{Pachucki2009}
K.~Pachucki, J.~Komasa, Nonadiabatic corrections to rovibrational levels of
  {H}$_{2}$, J. Chem. Phys. 130 (2009) 164113.

\bibitem{Piszczatowski2009}
K.~Piszczatowski, G.~\L{}ach, M.~Przybytek, J.~Komasa, K.~Pachucki,
  B.~Jeziorski, Theoretical determination of the dissociation energy of
  molecular hydrogen, J. Chem. Theory Comput. 5 (2009) 3039--3048.

\bibitem{Komasa2011}
J.~Komasa, K.~Piszczatowski, G.~\L{}ach, M.~Przybytek, B.~Jeziorski,
  K.~Pachucki, Quantum electrodynamics effects in rovibrational spectra of
  molecular hydrogen, J. Chem. Theory Comput. 7 (2011) 3105--3115.

\bibitem{Pachucki2010b}
K.~Pachucki, J.~Komasa, Rovibrational levels of {HD}, Phys. Chem. Chem. Phys.
  12 (2010) 9188--9196.

\bibitem{Wolniewicz1993}
L.~Wolniewicz, Relativistic energies of the ground state of the hydrogen
  molecule, \jcp 99 (1993) 1851--1868.

\bibitem{Wolniewicz1995}
L.~Wolniewicz, Nonadiabatic energies of the ground state of the hydrogen
  molecule, \jcp 103 (1995) 1792--1799.

\bibitem{Ramsey1952}
N.~F. Ramsey, Theory of molecular hydrogen and deuterium in magnetic fields,
  Phys. Rev. 85 (1952) 60--65.

\bibitem{Moss1993b}
R.~E. Moss, Calculations for vibration-rotation levels of {HD$^+$}, in
  particular for high {$N$}, Mol. Phys. 78 (1993) 371--405.

\bibitem{Korobov2006}
V.~I. Korobov, L.~Hilico, J.-P. Karr, Hyperfine structure in the hydrogen
  molecular ion, Phys. Rev. A 74 (2006) 040502.

\bibitem{Korobov2009}
V.~I. Korobov, L.~Hilico, J.-P. Karr, Relativistic corrections of
  m$\alpha^{6}$(m/{M}) order to the hyperfine structure of the {H}$_{2}^{+}$
  molecular ion, \pra 79 (2009) 012501.

\bibitem{Jefferts1968}
K.~B. Jefferts, Rotational hfs spectra of {H$_2^+$} molecular ions, Phys. Rev.
  Lett. 20 (1968) 39--41.

\bibitem{Jefferts1969}
K.~B. Jefferts, Hyperfine structure in the molecular ion {H$_2^+$}, Phys. Rev.
  Lett. 23 (1969) 1476--1478.

\bibitem{Korobov2015}
V.~I. Korobov, J.~C.~J. Koelemeij, L.~Hilico, J.-P. Karr, Test of the
  theoretical hyperfine structure of the molecular hydrogen ion at the 1-ppm
  level, arXiv:1510.05206 (2015).

\bibitem{Korobov2000}
V.~I. Korobov, Coulomb three-body bound-state problem: {V}ariational
  calculations of nonrelativistic energies, Phys. Rev. A 61 (2000) 064503.

\bibitem{Korobov2006a}
V.~I. Korobov, Leading-order relativistic and radiative corrections to the
  rovibrational spectrum of {H$_2^+$} and {HD$^+$} molecular ions, Phys. Rev. A
  74 (2006) 052506.

\bibitem{Korobov2014a}
V.~I. Korobov, L.~Hilico, J.-P. Karr, {m$\alpha^7$}-order corrections in the
  hydrogen molecular ions and antiprotonic helium, Phys. Rev. Lett. 112 (2014)
  103003.

\bibitem{Korobov2014b}
V.~I. Korobov, L.~Hilico, J.-P. Karr, Theoretical transition frequencies beyond
  0.1 ppb accuracy in {H$_2^{+}$}, {HD$^+$}, and antiprotonic helium, Phys.
  Rev. A 89 (2014) 032511.

\bibitem{Salumbides2013}
E.~J. Salumbides, J.~C.~J. Koelemeij, J.~Komasa, K.~Pachucki, K.~S.~E. Eikema,
  W.~Ubachs, Bounds on fifth forces from precision measurements on molecules,
  \prd 87 (2013) 112008.

\bibitem{Salumbides2014}
E.~Salumbides, W.~Ubachs, V.~Korobov, Bounds on fifth forces at the
  sub-{{\AA}}ngstr{\"{o}}m length scale, \jms 300 (2014) 65--69.

\bibitem{Salumbides2015b}
E.~J. Salumbides, A.~N. Schellekens, B.~Gato-Rivera, W.~Ubachs, Constraints on
  extra dimensions from precision molecular spectroscopy, New J. Phys. 17
  (2015) 033015.

\bibitem{Pohl2010}
R.~Pohl, A.~Antognini, F.~Nez, F.~D. Amaro, F.~Biraben, J.~M.~R. Cardoso, D.~S.
  Covita, A.~Dax, S.~Dhawan, L.~M.~P. Fernandes, A.~Giesen, T.~Graf, T.~W.
  H{\"{a}}nsch, P.~Indelicato, L.~Julien, C.~Y. Kao, P.~Knowles, E.-O.~L.
  Bigot, Y.-W. Liu, J.~A.~M. Lopes, L.~Ludhova, C.~M.~B. Monteiro,
  F.~Mulhauser, T.~Nebel, P.~Rabinowitz, J.~M.~F. dos Santos, L.~A. Schaller,
  K.~Schuhmann, C.~Schwob, D.~Taqqu, J.~F. C.~A. Veloso, F.~Kottmann, The size
  of the proton, Nature 466 (2010) 213--216.

\bibitem{Eides2001}
M.~Eides, H.~Grotch, V.~Shelyuto, Theory of light hydrogenlike atoms, Phys.
  Rep. 342 (2001) 63--261.

\bibitem{ArkaniHamed1998}
N.~{Arkani-Hamed}, S.~Dimopoulos, G.~Dvali, The hierarchy problem and new
  dimensions at a millimeter, \plb 429 (1998) 263--272.

\bibitem{Antoniadis1998}
I.~Antoniadis, N.~Arkani-Hamed, S.~Dimopoulos, G.~Dvali, New dimensions at a
  millimeter to a fermi and superstrings at a {TeV}, \plb 436 (1998) 257--263.

\bibitem{Klein1926}
O.~{Klein}, {Quantentheorie und f{\"u}nfdimensionale Relativit{\"a}tstheorie},
  Zeitschr. f. Phys. 37 (1926) 895--906.

\bibitem{ArkaniHamed1999}
N.~Arkani-Hamed, S.~Dimopoulos, G.~Dvali, Phenomenology, astrophysics, and
  cosmology of theories with submillimeter dimensions and {TeV} scale quantum
  gravity, Phys. Rev. D 59 (1999) 086004.

\bibitem{Witten1995}
E.~Witten, String theory dynamics in various dimensions, \npb 443 (1995)
  85--126.

\bibitem{Ubachs2015}
W.~{Ubachs}, J.~{Bagdonaite}, E.~J. {Salumbides}, M.~T. {Murphy}, L.~{Kaper},
  {Search for a drifting proton--electron mass ratio from {H$_2$}}, Submitted.

\bibitem{Balashev2014}
S.~A. {Balashev}, V.~V. {Klimenko}, A.~V. {Ivanchik}, D.~A. {Varshalovich},
  P.~{Petitjean}, P.~{Noterdaeme}, {Molecular hydrogen absorption systems in
  Sloan Digital Sky Survey}, \mnras 440 (2014) 225--239.

\bibitem{Pepe2010}
F.~A. {Pepe}, S.~{Cristiani}, R.~{Rebolo Lopez}, N.~C. {Santos}, A.~{Amorim},
  G.~{Avila}, W.~{Benz}, P.~{Bonifacio}, A.~{Cabral}, P.~{Carvas}, R.~{Cirami},
  J.~{Coelho}, M.~{Comari}, I.~{Coretti}, V.~{De Caprio}, H.~{Dekker},
  B.~{Delabre}, P.~{Di Marcantonio}, V.~{D'Odorico}, M.~{Fleury},
  R.~{Garc{\'{\i}}a}, J.~M. {Herreros Linares}, I.~{Hughes}, O.~{Iwert},
  J.~{Lima}, J.-L. {Lizon}, G.~{Lo Curto}, C.~{Lovis}, A.~{Manescau},
  C.~{Martins}, D.~{M{\'e}gevand}, A.~{Moitinho}, P.~{Molaro}, M.~{Monteiro},
  M.~{Monteiro}, L.~{Pasquini}, C.~{Mordasini}, D.~{Queloz}, J.~L. {Rasilla},
  J.~M. {Rebord{\~a}o}, S.~{Santana Tschudi}, P.~{Santin}, D.~{Sosnowska},
  P.~{Span{\`o}}, F.~{Tenegi}, S.~{Udry}, E.~{Vanzella}, M.~{Viel}, M.~R.
  {Zapatero Osorio}, F.~{Zerbi}, {ESPRESSO: the Echelle spectrograph for rocky
  exoplanets and stable spectroscopic observations}, in: SPIE Conf. Ser., Vol.
  7735, 2010, p.~9.

\bibitem{Murphy2007a}
M.~T. Murphy, T.~Udem, R.~Holzwarth, A.~Sizmann, L.~Pasquini, C.~Araujo-Hauck,
  H.~Dekker, S.~D'Odorico, M.~Fischer, T.~W. H\"{a}nsch, A.~Manescau,
  High-precision wavelength calibration of astronomical spectrographs with
  laser frequency combs, \mnras 380 (2007) 839--847.

\bibitem{Wilken2012}
T.~Wilken, G.~Lo~Curto, R.~A. Probst, T.~Steinmetz, A.~Manescau, L.~Pasquini,
  J.~I. Gonzalez~Hernandez, R.~Rebolo, T.~W. H\"{a}nsch, T.~Udem, R.~Holzwarth,
  {A spectrograph for exoplanet observations calibrated at the
  centimetre-per-second level}, Nature {485} ({2012}) 611--614.

\bibitem{Penzias1968}
A.~A. {Penzias}, K.~B. {Jefferts}, D.~F. {Dickinson}, A.~E. {Lilley},
  H.~{Penfield}, A search for line emission from singly ionized hydrogen
  molecules, \apj 154 (1968) 389--390.

\bibitem{Varshalovich1993a}
D.~A. Varshalovich, A.~V. Sannikov, {H$_2^+$} ions in the interstellar medium,
  Astron. Lett. 19 (2013) 290--293.

\bibitem{Shuter1986}
W.~L.~H. {Shuter}, D.~R.~W. {Williams}, S.~R. {Kulkarni}, C.~{Heiles}, {A
  search for vibrationally excited interstellar {H$_2^+$}}, \apj 306 (1986)
  255--258.

\bibitem{Dewdney2009}
P.~E. Dewdney, P.~J. Hall, R.~T. Schillizi, T.~J. L.~W. Lazio, The {S}quare
  {K}ilometer {A}rray, Proc. IEEE 97 (2009) 1482--1496.

\bibitem{Teller1930}
E.~Teller, {\"{U}}ber das {W}asserstoffmolek{\"{u}}lion, Zeitschr. f. Phys. 61
  (1930) 458--480.

\bibitem{Carrington1989}
A.~Carrington, I.~R. McNab, C.~A. Montgomerie, Microwave electronic spectrum of
  the {H$_2^+$} ion, \cpl 160 (1989) 237--242.

\bibitem{Carrington1988b}
A.~Carrington, I.~R. McNab, C.~A. Montgomerie, Vibration-rotation spectroscopy
  of the {HD$^+$} ion near the dissociation limit, \molp 64 (1988) 983--995.

\bibitem{Heitler1927}
W.~Heitler, F.~London, Wechselwirkung neutraler {A}tome und homopolare
  {B}indung nach der {Q}uantenmechanik, Zeitschr. f. Phys. 44 (1927) 455--472.

\bibitem{Sugiura1927}
Y.~Sugiura, {\"{U}}ber die {E}igenschaften des {W}asserstoffmolek{\"{u}}ls im
  {G}rundzustande, Zeitschr. f. Phys. A 45 (1927) 484--492.

\bibitem{Wang1928}
S.~C. Wang, The problem of the normal hydrogen molecule in the new quantum
  mechanics, Phys. Rev. 31 (1928) 579--586.

\bibitem{Rosen1931}
N.~Rosen, The normal state of the hydrogen molecule, Phys. Rev. 38 (1931)
  2099--2114.

\bibitem{James1933}
H.~M. James, A.~S. Coolidge, The ground state of the hydrogen molecule, \jcp 1
  (1933) 825--835.

\bibitem{Kolos1960}
W.~Ko{\l}os, C.~C.~J. Roothaan, Accurate electronic wave functions for the
  {H$_2$} molecule, Rev. Mod. Phys. 32 (1960) 219--232.

\bibitem{Kolos1962}
W.~Ko{\l}os, L.~Wolniewicz, A complete non-relativistic treatment of the
  {H$_2$} molecule, Phys. Lett. 2 (1962) 222--223.

\bibitem{Kolos1964}
W.~Ko\l{}os, L.~Wolniewicz, Accurate adiabatic treatment of the ground state of
  the hydrogen molecule, J. Chem. Phys. 41 (1964) 3663--3673.

\bibitem{Wolniewicz1966}
L.~Wolniewicz, Vibrational{-}rotational study of the electronic ground state of
  the hydrogen molecule, \jcp 45 (1966) 515--523.

\bibitem{Kolos1968}
W.~Ko\l{}os, L.~Wolniewicz, Improved theoretical ground-state energy of the
  hydrogen molecule, J. Chem. Phys. 49 (1968) 404--410.

\bibitem{Bunker1972}
P.~R. Bunker, On the breakdown of the {B}orn{-O}ppenheimer approximation for a
  diatomic molecule, \jms 42 (1972) 478--494.

\bibitem{Kolos1975}
W.~Ko\l{}os, L.~Wolniewicz, Improved potential energy curve and vibrational
  energies for the electronic ground state of the hydrogen molecule, J. Mol.
  Spectrosc. 54 (1975) 303--311.

\bibitem{Bishop1978}
D.~M. Bishop, L.~M. Cheung, Rigorous theoretical investigation of the ground
  state of {H$_2$}, Phys. Rev. A 18 (1978) 1846--1852.

\bibitem{Wolniewicz1983}
L.~Wolniewicz, The {X$^1\Sigma_g^+$} state vibration{-}rotational energies of
  the {H$_2$}, {HD}, and {D$_2$} molecules, \jcp 78 (1983) 6173--6181.

\bibitem{Kolos1986}
W.~Ko\l{}os, K.~Szalewicz, H.~Monkhorst, New {B}orn{-}{O}ppenheimer potential
  energy curve and vibrational energies for the electronic ground state of the
  hydrogen molecule, J. Chem. Phys. 84 (1986) 3278--3283.

\bibitem{Schwartz1987}
C.~Schwartz, R.~J. Le~Roy, Nonadiabatic eigenvalues and adiabatic matrix
  elements for all isotopes of diatomic hydrogen, \jms 121 (1987) 420--439.

\bibitem{Kolos1993}
W.~Ko\l{}os, J.~Rychlewski, Improved theoretical dissociation energy and
  ionization potential for the ground state of the hydrogen molecule, \jcp 98
  (1993) 3960--3967.

\bibitem{Witmer1926}
E.~E. Witmer, Critical potentials and the heat of dissociation of hydrogen as
  determined from its ultra-violet band spectrum, Phys. Rev. 28 (1926)
  1223--1241.

\bibitem{Dieke1927}
G.~H. Dieke, J.~J. Hopfield, The structure of the ultra-violet spectrum of the
  hydrogen molecule, Phys. Rev. 30 (1927) 400--417.

\bibitem{Richardson1929}
O.~W. Richardson, P.~M. Davidson, The spectrum of {H$_2$} - {T}he bands
  analogous to the parahelium line spectrum - {P}art {II}, Proc. Roy. Soc.
  London Ser. A 123 (1929) 466--488.

\bibitem{Beutler1934}
H.~Beutler, A.~Deubner, H.~Junger, Die {D}issoziationswarme des
  {W}asserstoffmolekuls {H$_2$}, aus einem neuen ultravioletten
  {R}esonanzbandenzug bestimmt, Zeitschr. f. Phys. Chem. 27B (1934) 287--302.

\bibitem{Beutler1935}
H.~Beutler, The dissociation heats of the hydrogen molecule {H$_2$}, determined
  at 850 angstrom from the rotation structure on the long-wave border of the
  absorption continuum, Zeitschr. f. Phys. Chem. 29B (1935) 315--327.

\bibitem{Herzberg1961}
G.~Herzberg, A.~Monfils, The dissociation energies of the {H$_2$}, {HD}, and
  {D$_2$} molecules, \jms 5 (1961) 482--498.

\bibitem{Herzberg1970}
G.~Herzberg, The dissociation energy of the hydrogen molecule, J. Mol.
  Spectrosc. 33 (1970) 147--168.

\bibitem{Stwalley1970}
W.~C. Stwalley, The dissociation energy of the hydrogen molecule using
  long-range forces, \cpl 6 (1970) 241--244.

\bibitem{Balakrishnan1992}
A.~Balakrishnan, V.~Smith, B.~P. Stoicheff, Dissociation energy of the hydrogen
  molecule, \prl 68 (1992) 2149--2152.

\bibitem{Eyler1993}
E.~E. Eyler, N.~Melikechi, Near-threshold continuum structure and the
  dissociation energies of {H}$_{2}$, {HD}, and {D}$_{2}$, Phys. Rev. A 48
  (1993) R18.

\bibitem{Zhang2004}
Y.~P. Zhang, C.~H. Cheng, J.~T. Kim, J.~Stanojevic, E.~E. Eyler, Dissociation
  energies of molecular hydrogen and the hydrogen molecular ion, \prl 92 (2004)
  203003.

\bibitem{Sprecher2011}
D.~Sprecher, C.~Jungen, W.~Ubachs, F.~Merkt, Towards measuring the ionisation
  and dissociation energies of molecular hydrogen with sub-{MHz} accuracy,
  Farad. Discuss. 150 (2011) 51--70.

\bibitem{Sprecher2013}
D.~Sprecher, M.~Beyer, F.~Merkt, Precision measurement of the ionisation energy
  of the {$3d\sigma$} {GK} state of {H$_2$}, \molp 111 (2013) 2100--2107.

\bibitem{Zhang2008}
H.~Zhang, G.~Wang, L.~Guo, A.~Geng, Y.~Bo, D.~Cui, Z.~Xu, R.~Li, Y.~Zhu,
  X.~Wang, C.~Chen, 175 to 210 nm widely tunable deep-ultraviolet light
  generation based on {KBBF} crystal, \apb 93 (2008) 323--326.

\bibitem{Morgenweg2014}
J.~Morgenweg, I.~Barmes, K.~S.~E. Eikema, Ramsey-comb spectroscopy with intense
  ultrashort laser pulses, \nph 10 (2014) 30--33.

\bibitem{Morgenweg2014b}
J.~Morgenweg, K.~S.~E. Eikema, Ramsey-comb spectroscopy: {T}heory and signal
  analysis, \pra 89 (2014) 052510.

\bibitem{Pachucki2014}
K.~Pachucki, J.~Komasa, Accurate adiabatic correction in the hydrogen molecule,
  J. Chem. Phys. 141 (2014) 224103.

\bibitem{Pachucki2015}
K.~Pachucki, J.~Komasa, Leading order nonadiabatic corrections to rovibrational
  levels of {H$_2$}, {D$_2$}, and {T$_2$}, \jcp 143 (2015) 034111.

\bibitem{Pachucki2011}
K.~Pachucki, J.~Komasa, Gerade-ungerade mixing in the hydrogen molecule, \pra
  83 (2011) 042510.

\bibitem{Veirs1987}
D.~Veirs, G.~Rosenblatt, Raman line positions in molecular hydrogen: {H$_2$},
  {HD}, {HT}, {D$_2$}, {DT}, and {T$_2$}, \jms 121 (1987) 401--419.

\bibitem{Chuang1987}
M.-C. Chuang, R.~N. Zare, Rotation-vibration spectrum of {HT}: {L}ine position
  measurements of the 1-0, 4-0, and 5-0 bands, \jms 121 (1987) 380--400.

\bibitem{Black1976}
J.~H. {Black}, A.~{Dalgarno}, {Interstellar {H$_2$} - {T}he population of
  excited rotational states and the infrared response to ultraviolet
  radiation}, \apj 203 (1976) 132--142.

\bibitem{Pachucki2011a}
K.~Pachucki, J.~Komasa, Magnetic dipole transitions in the hydrogen molecule,
  \pra 83 (2011) 032501.

\bibitem{Hogan2009}
S.~D. Hogan, C.~Seiler, F.~Merkt, Rydberg-state-enabled deceleration and
  trapping of cold molecules, Phys. Rev. Lett. 103 (2009) 123001.

\bibitem{Seiler2011}
C.~Seiler, S.~D. Hogan, F.~Merkt, Trapping cold molecular hydrogen, Phys. Chem.
  Chem. Phys. 13 (2011) 19000--19012.

\bibitem{Schiller2014}
S.~Schiller, D.~Bakalov, V.~I. Korobov, Simplest molecules as candidates for
  precise optical clocks, Phys. Rev. Lett. 113 (2014) 023004.

\bibitem{Wing1976}
W.~H. Wing, G.~A. Ruff, W.~E. Lamb, J.~J. Spezeski, Observation of the infrared
  spectrum of the hydrogen molecular ion {HD$^+$}, Phys. Rev. Lett. 36 (1976)
  1488--1491.

\bibitem{Karr2005}
J.~P. Karr, S.~Kilic, L.~Hilico, Energy levels and two-photon transition
  probabilities in the {HD$^+$} ion, \jpb 38 (2005) 853--866.

\bibitem{Tran2013}
V.~Q. Tran, J.-P. Karr, A.~Douillet, J.~C.~J. Koelemeij, L.~Hilico, Two-photon
  spectroscopy of trapped {HD}${}^{+}$ ions in the {L}amb-{D}icke regime, Phys.
  Rev. A 88 (2013) 033421.

\end{thebibliography}

\end{document}